\documentclass[AMA,STIX1COL]{WileyNJD-v2}

\articletype{Article Type}

\received{26 April 2016}
\revised{6 June 2016}
\accepted{6 June 2016}

\let\bibhang\relax
\let\bibfont\relax

\let\href\relax

\usepackage{csquotes} 
\usepackage{hyperref}
\usepackage{ragged2e} 
\usepackage{caption} 
\usepackage{float} 
\usepackage{tabularx}
\usepackage{tikz}
\usepackage{listings}
\usepackage{algpseudocode}
\usepackage{amsmath}
\usepackage{fancyvrb}
\usepackage{xcolor}
\usetikzlibrary{arrows, positioning}
\usetikzlibrary{shapes,arrows}

\DeclareUnicodeCharacter{2028}{-}
\usepackage{pgfplots}
\pgfplotsset{compat=1.18}
\usepackage{pgf-pie}
\usepackage{subcaption}
\usepackage{afterpage}
\usepackage{pgfplotstable}
\usepackage{hyperref}

\usepackage[backend=biber,style=numeric,sorting=none]{biblatex}
\begin{document}

\title{Best Practices for Developing Computational and Data-Intensive (CDI) Applications}

\author[1]{Parinaz Barakhshan* (0000-0001-7232-3923)}

\author[2]{Rudolf Eigenmann (0000-0003-1651-827X)}

\authormark{Parinaz Barakhshan \& Rudolf Eigenmann }

\address[1]{\orgdiv{Department of Electrical and Computer Engineering}, \orgname{University of Delaware}, \orgaddress{\state{Delaware}, \country{USA}}}

\address[2]{\orgdiv{Department of Electrical and Computer Engineering}, \orgname{University of Delaware}, \orgaddress{\state{Delaware}, \country{USA}}}

\corres{*Parinaz Barakhshan,\email{parinazb@udel.edu}}

\presentaddress{University of Delaware, Evans Hall,
139 The Green, Newark, DE 19716}

\abstract[Summary]{
High-quality computational and data-intensive (CDI) applications are critical for advancing research frontiers in almost all disciplines. Despite their importance, there is a significant gap due to the lack of comprehensive best practices for developing such applications. CDI projects, characterized by specialized computational needs, high data volumes, and the necessity for cross-disciplinary collaboration, often involve intricate scientific software engineering processes. The interdisciplinary nature necessitates collaboration between domain scientists and CDI professionals (Xperts), who may come from diverse backgrounds. 

This paper aims to close the above gap by describing practices specifically applicable to CDI applications. They include general software engineering practices to the extent that they exhibit substantial differences from those already described in the literature as well as practices that have been called pivotal by Xperts in the field.

The practices were evaluated using three main metrics: (1) participants' experience with each practice, (2) their perceived impact, and (3) their ease of application during development. The evaluations involved participants with varying levels of experience in adopting these practices. Despite differing experience levels, the evaluation results consistently showed high impact and usability for all practices.

By establishing a best-practices guide for CDI research, the ultimate aim of this paper is to enhance CDI software quality, improve approaches to computational and data-intensive challenges, foster interdisciplinary collaboration, and thus accelerate scientific innovation and discovery.
}


\maketitle

\section{Introduction}
\label{intro}
Computational and data-intensive (CDI) applications are critical in modern research for their ability to process and analyze vast amounts of data, as well as execute complex computational tasks to address scientific problems. CDI applications cater to the demands of data-driven and computation-heavy research projects, requiring not only accuracy but also high levels of efficiency and scalability. 
This paper aims to increase the productivity of developers of such applications by describing and evaluating a set of best practices, gathered from a diverse pool of researchers.

Best practices are guidelines or standards that outline the most effective and efficient approaches to achieving desired outcomes in application development. They provide a shared knowledge base and a common framework for practitioners, researchers, and even experts to build upon. Although general software engineering has a well-defined set of best practices for developers, which has applicability to CDI research software, a specific guide is needed for the following reasons.

CDI research distinguishes itself from general software engineering through specific characteristics: (1) CDI problem domains often involve applying computational methods to domain-specific challenges across various scientific disciplines. (2) Interdisciplinary collaboration is often required, bringing together diverse expertise to address complex problems. (3) CDI research is inherently data-intensive, dealing with massive data volumes to extract insights. (4) Significant computational resources are required for complex computations and data processing.

Another key difference between CDI research and general software engineering practices lies in the imperative for collaboration among domain scientists and CDI professionals, emphasizing team dynamics beyond mere software development. This is due to the interdisciplinary nature of such projects, which requires collaboration between domain scientists and CDI support professionals —- referred to as Xperts —- to tackle computer engineering challenges. Such collaboration is less common in traditional software engineering, where the focus is primarily on software development and engineering practices. In this article, we use the terms Xperts, research software engineers (RSEs), research facilitators, and research programmers interchangeably to denote professionals who support CDI research software development.

Xperts come from a wide range of disciplines that extend beyond computer science. Many are domain scientists who have delved into CDI application development, eager to learn about new scientific fields as they support their colleagues. Alternatively, some have developed their expertise through practical experience, making significant contributions to CDI projects. Irrespective of their origin, Xperts are expected to tackle computer engineering issues in CDI research projects. They are instrumental in working alongside domain scientists, lending their technical know-how, and navigating the unique computational and data challenges of CDI applications.

The main goal of this article is to introduce a set of best practices for developing computational and data-intensive (CDI) applications, aiming to boost efficiency, effectiveness, and innovation in collaboration with science teams. By promoting the adoption of these best practices that are tailored to the characteristics of CDI research, CDI support professionals or developers can enhance the development and optimization of CDI applications and improve their efficiency and effectiveness. 

The best practices highlighted in this article encompass newly introduced practices tailored specifically for CDI applications, those substantially diverging from conventional software engineering (SE) practices, or those emphasized for their importance by the Xpert Network community~\cite{Xpert, XpertNework22}.

Several efforts contributed to identifying and evaluating best practices, as described in Section~\ref{efforts}:
(i) Through the Xpert Network~\cite{XpertNework22, xpert2019}, we collected insights and best practices from CDI researchers, tool developers, and Xperts in the CDI field. 
(ii) In the Atom project~\cite{Atom, atomarxiv2022portal, AtomGateways21, AtomGateways22}, we implemented and evaluated the practices identified via the Xpert network within a specific CDI project. This effort also revealed additional practices and insights. 
(iii) Through surveys we assessed the impact and usability of the practices identified from these two sources. The responses also pointed to associated challenges, benefits, limitations, and supporting tools.

The identified best practices are explained in Section~\ref{BestPractices}. Each practice is described in terms of a highlight, purpose, impact, recommended integration method into the development process, significance in the context of CDI development, challenges and limitations of implementation, as well as supporting tools and resources.

Subsequently, Section~\ref{evaluation} elaborates on our approach to evaluating CDI best practices through three distinct surveys that contributed unique insights. 
(i) Surveys targeted at CDI Researchers: We collected feedback from a diverse group of researchers and practitioners across various disciplines, involving 46 participants utilizing HPC resources at the University of Delaware (UD). This survey provided a broad perspective on how these best practices are adopted and perceived in the research community.
(ii) Case study evaluation via the Atom Project: By applying the practices identified by the Xpert Network in the Atom Project, we could observe their impact and usability. This effort also led to the identification of additional practices. Feedback from seven project participants, including CDI professionals and domain scientists, informed our understanding of how these practices perform in a real-world setting, highlighting their practical benefits and challenges.
(iii) Expert Reviews: We engaged 14 CDI support professionals, or Xperts, who were actively engaged in Xpert Network activities, to obtain specialized assessments and insights. Their significant experience with CDI applications allowed for a detailed and expert analysis.

Our work is related to several key initiatives in the field of scientific software development that have made significant contributions to improving the quality and reliability of computational and data-intensive research across various domains. Section~\ref{relatedWork} discusses these contributions. Notable efforts include Better Scientific Software (BSSw)~\cite{bssw}, the Research Software Engineering (RSE) community~\cite{USRSE}, ACCESS (Advanced Cyberinfrastructure Coordination Ecosystem: Services \& Support)~\cite{ACCESS, ACCESSweb}, and MolSSI (Molecular Sciences Software Institute)~\cite{MOLSII}. 
Our effort distinguishes itself by providing and evaluating a set of best practices tailored specifically for CDI support professionals and developers. These guidelines address their unique requirements for working in collaboration with science teams and developing CDI applications. 

Key contributions of our research include the identification of seventeen best practices specific to computational and data-intensive (CDI) research, along with providing practical insights on challenges and limitations tied to each practice within the CDI research landscape, and an overview of tools commonly utilized in CDI projects to efficiently implement these practices. 

\section{How we collected our information: Identifying Best Practices}
\label{efforts}
The information presented in this paper was gathered through three main sources: the Xpert Network~\cite{XpertNework22}, the Atom Project case study~\cite{AtomGateways21, AtomGateways22, atomarxiv2022portal}, and surveys that engaged a broad audience of researchers and practitioners. These efforts served to identify relevant best practices and provide data for their evaluation.

\subsection{Xpert Network}
The key initiative in this process was the establishment of the Xpert Network, a collaborative platform catering to: (i) researchers involved in the development and utilization of CDI applications, (ii) Xperts ( Facilitators~\cite{neeman2019virtual}, or Research Software Engineers (RSEs)~\cite{USRSE, rse8994167}), professionals providing support with CDI technology and methodologies,
(iii) domain experts and scientists from universities and research institutions,
and (iv) developers creating tools to facilitate the creation and usage of CDI applications.
Through webinars, workshops, and Birds-of-a-Feather (BoF) sessions, this diverse community shared knowledge, fostered collaboration, and identified best practices and essential tools for developing CDI applications~\cite{XpertNework22, 2021exchanging,xpert2019, Xpert}. 

Participants of the Xpert Network activities represent over 60 distinct projects, universities, national labs, and international institutions. Notable projects include ACCESS (Advanced Cyberinfrastructure Coordination Ecosystem: Services \& Support)~\cite{ACCESS}, CyVerse~\cite{cyverse}, MolSSI (Molecular Sciences Software Institute)~\cite{MOLSII}, and US-RSE (United States Research Software Engineer Association)~\cite{USRSE}. These projects cover a broad spectrum of disciplines and expertise, fostering collaboration and knowledge exchange among researchers, domain experts, and computational professionals. More information on participating resources and institutions is available on the project home page~\cite{xpertcdi}.

The collective expertise and varied experiences of participants in the Xpert Network were instrumental in formulating the practices discussed later in Section \ref{BestPractices}. This effort resulted in the formulation of the first 15 practices.

\subsection{Atom Project Case Study}\label{sec:atom}
We used a CDI project in which we were directly involved, the {\it Atom Project}~\cite{Atom}, for gauging the applicability and completeness of the identified practices. 
The Atom Project is a collaboration of physicists and computer scientists to develop computational applications and a community web portal for calculating and presenting data about certain properties of atoms and ions \cite{AtomGateways21, AtomGateways22, atomarxiv2022portal}. Examples of such properties are {\it transition rates} and {\it polarizabilities}. Among the software development tasks were the creation of the computational applications and the web portal server and client, including a database of the requisite physics data.

Through this collaboration, the unique challenges and complexities of interdisciplinary collaborations were highlighted. The effort led to the identification of two additional best practices, described as the final two in Section~\ref{BestPractices}. The Atom Project participants also contributed to the surveys used in Section~\ref{evaluation}.

\subsection{Surveys}
To enable the evaluation of the identified best practices in Section~\ref{evaluation}, the primary means was the creation of surveys. We devised and circulated surveys tailored to various participants, each serving distinct purposes. These purposes include the validation of the practices, the assessment of their impact and usability, as well as the understanding of practical applications and challenges. We also gathered information about tools that support the practices. We targeted three different groups with the surveys, including general CDI researchers, Xperts (CDI support professionals), and the participants of the Atom project case study.

\section{Identified Best Practices}
\label{BestPractices}
The following 17 subsections present each of the identified best practices in terms of 
(i) a one-sentence recommendation,
(ii) a brief description of the practice,
(iii) the expected impact of the practice,
(iv) recommended methods for integrating the practice into the development process, 
(v) an explanation of how the practice differs from those found in general software engineering,
(vi) anticipated challenges and limitations of employing the practice, and
(vii) a list of tools and resources that help support the practice.

The first five practices enable interdisciplinary collaboration. They are specific to the  CDI context and rarely found in general software engineering recommendations. The remaining 12 practices are software development oriented but differentiate themselves from general software engineering practices through distinct features or particular emphasis given by our Xpert Network participants. Paragraphs {\it ``Difference from General Software Engineering Practices''} elaborate this point for each practice.

\subsection{ BP1 - Onboarding Xperts from Diverse Backgrounds}
\label{bp:onboarding}
\begin{center}
  \fbox{\begin{minipage}{0.98\textwidth}
  Devise an onboarding process that is tailored to the specific backgrounds of individual Xperts.
  \end{minipage}}
\end{center}

\paragraph{The Practice} Xperts bring diverse backgrounds and a wide range of expertise to the team. Members of Xpert groups can come from various fields, including domain sciences and computer science, or they may have gained experience supporting CDI science teams through on-the-job training. While individuals with a computer science background often have a natural grasp of Unix commands, parallel programming models, and version control, these very skills may be critical best practices for Xperts with a domain-science background to acquire. On the other hand, Xperts who have a domain degree are often well acquainted with overcoming the terminology gap discussed in Section~\ref{bp:terminologygap}, which greatly facilitates communication with new science collaborators. 
Training programs for onboarding Xperts must take these differences into account, allowing trainees to focus on unfamiliar best practices. 

\paragraph{Impact} 
The main impact is a shortened onboarding process, yielding a productive team member quickly. 

\paragraph{Recommended Integration Methods}
\begin{itemize}
    \item Conduct a thorough assessment of each Xpert's knowledge, skills, and background to identify their specific training needs. 
    \item Create custom training modules for specific skills and practices. Combine the modules as suited for the training needs of individuals.
\end{itemize}

\paragraph{Difference from General Software Engineering Practices}
While onboarding is a standard practice in general software engineering, a key distinction in onboarding CDI research application developers arises from the interdisciplinary nature of CDI projects and the diversity of Xpert backgrounds.

\paragraph{Challenge(s) \& Limitation(s) }
Personalized onboarding as opposed to general onboarding, is initially more expensive and resource-intensive. However, the long-term gains of increased productivity and team performance validate the initial investment. 

\paragraph{Supportive Tools and Resources} 
\label{collaborationTools}
The Xpert Network discussions have pointed to a number of tools that assist with general onboarding. They include Slack~\cite{slack}, Zoom~\cite{zoom}, Microsoft Teams~\cite{microsoftteams}, and Google Workspace~\cite{googleworkspace} for team communication, file sharing, and project collaboration.

\subsection{BP2 - Understanding the Academic Environment}
\label{sec:academic-env}
\label{bp:academic}
\begin{center}
  \fbox{\begin{minipage}{0.98\textwidth}
    Familiarize Xperts with the academic reward system, research priorities, and side activities that strengthen the science team's academic standing.
  \end{minipage}}
\end{center}

\paragraph{The Practice} An issue for Xperts that is essentially absent in a general software engineering context is the importance of understanding the academic environment in order to navigate the unique dynamics and expectations of science teams. This issue is especially relevant for software engineers with backgrounds in industry, where hierarchical organization and the overriding goal of creating a reliable product as rapidly as possible are the norm. Understanding the academic reward system, research priorities, and the many side activities that researchers may get engaged in to maintain the academic standing of the science team will influence project decisions. For example, the need to be innovative might take priority over delivering production-quality applications; or publishing a research paper may compete with creating periodic project reports.  
In fact, some of the best practices need to be understood from this viewpoint, such as the issues of documentation and testing, mentioned in Sections~\ref{bp:documentation} and~\ref{bp:testing}. 
Another distinction of the academic environment is its dynamic nature, where project specifications may change "overnight", driven by the unpredictability of scientific innovation. Section~\ref{bp:domainproblem} on understanding the domain problem and developing a project plan will expand on this issue.

\paragraph{Impact} 
Comprehending academic factors will directly affect the setting of realistic goals, expectations, and timelines. It will also improve productive collaboration between Xperts and domain researchers.

\paragraph{Recommended Integration Methods} 
\begin{itemize}
 \item Devise an onboarding module that introduces new Xperts to the local academic environment. Describe expectations and job performance criteria, such as (co-)authoring scientific publications and grant writing, general goals of science teams to be assisted, prototyping versus production-quality code development, and the time devoted to tasks beyond software development, such as documentation, research, presentations, mentoring, and training.
 \item At the beginning of each new project, review these expectations and criteria jointly with the domain science teams that will be supported.
\end{itemize}

\paragraph{Difference from General Software Engineering Practices}
This practice is essentially absent in a general software engineering context. 

\paragraph{Challenge(s) \& Limitation(s) } 
The diversity of academic environments and research disciplines that Xperts support makes it difficult to define general recommendations. It requires that expectations be carefully defined for each team and project. An ability of Xperts to be sensitive to and adapt to new environments is critical.

\paragraph{Supportive Tools and Resources }
There are no specific resources recommended for this practice. 

\subsection{BP3 - Developing a Breadth of Skills for the Effective Handling of Projects}
\label{bp:breadth}
\begin{center}
  \fbox{\begin{minipage}{0.98\textwidth}
    Build partnerships with external Xperts to prepare for the diversity of supported domain projects and their required skill sets.
  \end{minipage}}
\end{center}
\paragraph{The Practice} 
Modern CDI applications encompass a wide spectrum of technologies, including computing paradigms, programming languages, architectures, and algorithms. With the continuous evolution of CDI applications and their diverse technological requirements, it becomes challenging for an individual or a small team of Xperts to possess all essential skills. For instance, emerging CDI applications might necessitate expertise in machine learning techniques, an area that individual Xperts or small teams serving a vast CDI research community may struggle to fully cover. Furthermore, these professionals often engage in tasks spanning the entire software life cycle and contribute to project management. Hence, it becomes imperative to identify instances where seeking external advice is necessary. By establishing and nurturing connections with other Xpert teams, individuals and teams can ensure access to supplementary skills and knowledge that go beyond their current expertise.

\paragraph{Impact} 
Applying this practice and maintaining contacts with other expert teams enables Xperts to enhance their readiness in handling the multitude of emerging tasks throughout the life cycle of a CDI application and can effectively navigate the complexities of the application development process, ensuring successful project outcomes. 

\paragraph{Recommended Integration Methods}
\begin{itemize}
    \item Participate in communities of practice where Xperts with similar interests or expertise share insights, address challenges, and collaborate on solutions. An illustrative example is the Xpert Network~\cite{XpertNework22}, offering a dedicated space for Xperts to connect, exchange experiences, discuss challenges and share ideas. These interactions foster cross-team collaboration and the exchange of valuable lessons learned.
    \item Attend networking events, conferences, workshops, and seminars in the CDI field to connect with professionals from diverse backgrounds and engage in knowledge-sharing opportunities.
    \item Join online forums, social media groups, and professional networks related to the CDI domain to actively participate in discussions, share insights, and connect with Xperts from various projects.
    \item Consider outsourcing projects to Xpert teams with expertise in skills not available locally.
\end{itemize}

\paragraph{Difference from General Software Engineering Practices}
This practice is rarely found in general software engineering guides. We attribute this to the smaller average size of academic research teams compared to industry teams, making it more difficult to cover all the necessary skills. 
Also, research is intrinsically exploratory, including the test of novel technologies, thus expanding the range of skills Xpert groups need to be aware of. Finally, intellectual property is guarded more closely in industry settings, whereas seeking and providing help across groups and institutions is often done freely in academic environments.

\paragraph{Challenge(s) \& Limitation(s)}
Limited resources, in terms of time, funding, and personnel, are a primary challenge to building an Xpert team that covers the required skills and keeps up with new technologies. This is a hard constraint. However, there is also a cultural aspect of interdisciplinary collaboration and knowledge sharing among Xpert teams and within the academic community, which can be addressed with proper project management: Team members must be encouraged and given the time to seek interactions with related groups, participate in professional discussion forums and events, and continue their education.

\paragraph{Supportive Tools and Resources}
While the practice is more reliant on personal connections, networking, and maintaining contacts with other Xpert teams, there are a few tools and resources that can support this practice in CDI development. 
\begin{itemize}
    \item Professional networks and communities, such as LinkedIn~\cite{linkedin}, Twitter~\cite{twitter}, Xpert Network~\cite{XpertNework22}, and research forums, provide opportunities to connect with peers and address open problems within the field. 
    \item Online forums and discussion boards provide platforms for connecting with a wider community and learning from shared experiences. An example is Connect Cyberinfrastructure (cnct.ci~\cite{Connect.CI}), which offers access to various portals serving different segments of the research computing and data community. Users can join and actively participate in threads to engage with others and gain valuable insights.
\end{itemize} 

\subsection{BP4 - Collaborative Assistance Between Xperts and Domain Scientists}
\label{bp:assistance}
\begin{center}
  \fbox{\begin{minipage}{0.98\textwidth}
     Assist domain scientists through short-term, close collaborations, rather than from behind a service desk.
  \end{minipage}}
\end{center}
\paragraph{The Practice} A recommended form of interaction between Xperts and CDI domain scientists is through {\em collaborative assistance}. For a period of one to several months, Xperts work side-by-side (physically or virtually) with the domain scientists whose projects they support. 
While collaboratively working on the problem, each team member focuses on their respective areas of expertise, with Xperts handling computer-engineering issues and domain scientists addressing application science problems. This division of labor allows for immediate problem-solving and efficient utilization of skills. It also reduces the need for cross-training. 
Over the course of the joint work, the collaborators tend to pick up each others' knowledge, skills, and terminology, allowing the domain researchers to continue the project independently, or with minimal remote support, after the collaborative assistance period. This collaborative model was initially pioneered by XSEDE's ECSS group~\cite{ECSS} and has since been successfully implemented by other Xpert teams~\cite{xpert2019}.

\paragraph{Impact} Through collaborative assistance, Xperts and domain scientists leverage each other's skills and expertise, leading to accelerated progress and development of a higher-quality end product. This collaboration also facilitates the exchange of knowledge and the mutual understanding of each other's skills and terminologies. Surveys by the XSEDE ECSS group~\cite{ECSS} have found that collaborative models, where domain scientists work with Xperts, can significantly enhance efficiency~\cite{6866038, romanus2012anatomy}.

\paragraph{Recommended Integration Methods}
\begin{itemize}
\item Allocating projects to Xpert resources is a key step. Carefully match the skills needed for the project with those available among the Xperts. A competitive proposal process may be indicated if requested projects exceed available resources.
\item The initial project phase is especially critical; in-person collaboration was highly recommended by Xpert Network participants. In addition to the issues of Section~\ref{sec:academic-env}, devote sufficient time to the discussion of overall domain science goals, specific collaboration goals, terminology, possible approaches, project plan, and division of work between Xperts and domain scientists.
\item Xperts and their home teams need to keep in regular contact, such as weekly meetings to brief the home team on progress and discuss open problems. This is especially important for in-person collaborations at remote sites.
\item Be sure to plan your time into the schedule for professional development, even when at remote sites.
\end{itemize}

\paragraph{Difference from General Software Engineering Practices}
While this practice may also be applicable in an industrial setting, it has not been covered in the general software engineering literature.

\paragraph{Challenge(s) \& Limitation(s)}
Adding a competitive process introduces overhead for setup, proposal writing, and evaluation; simplified allocations, FIFO processing, and short collaborations may be alternatives. Travel logistics and space constraints for in-person collaborations are obvious challenges, with recent advances in remote participation providing reasonable options. The hand-over from Xpert to domain scientist at the end of the collaborative project can be an issue; a schedule of weekly (or initially even daily) brief ``check-in Zoom calls'', with decreasing frequency is advisable. 

\paragraph{Supportive Tools and Resources}
Among the supporting tools mentioned by the participants are collaboration platforms for information sharing, virtual meeting tools for real-time interactions, as discussed in Section~\ref{collaborationTools}, and version control systems for collaborative code management, which will be detailed in Section~\ref{bp:sourcecode}.

\subsection{BP5 - Overcoming the Terminology Gap Between Computer and Domain Sciences}
\label{bp:terminologygap}
\begin{center}
  \fbox{\begin{minipage}{0.98\textwidth}Avoid domain-specific language, define key terminologies, identify and clarify shared terminologies. 

  \end{minipage}}
\end{center}

\paragraph{The Practice}
The gap between computer science and domain science lingo is an often-mentioned issue. The challenge can be big if the same term is used by both collaborators but with different meanings. Xpert Network participants reported significant confusion and even incorrect project executions due to this challenge. Awareness of the issue and patience in trying to understand the collaborators' viewpoints is critical. Keeping the vocabulary to the essentials and investing time in explaining new terms is key to successful collaboration. Using many examples and frequent feedback from both sides will help bridge this gap. 
\paragraph{Impact}
Overcoming terminology gaps reduces misunderstandings and errors in project development, ensuring that the intended goals and objectives are accurately translated and achieved. It brings clarity to reports and research papers, and it facilitates the transfer of knowledge between computer science and domain sciences.

\paragraph{Recommended Integration Methods} 
\begin{itemize}
    \item Encourage team members to look out for terminology gaps and resolve them through frequent communication, especially at project begin.
    \item Explain terms and concepts using simple language and examples. 
    \item Develop a glossary of terms, also serving for project documentation. 
\end{itemize}

\paragraph{Difference from General Software Engineering Practices}
The terminology gap is wider in CDI research contexts compared to general software engineering. While understanding the client is important for all software engineers, Xperts in CDI research contexts may need to become familiar with highly specialized domain vocabularies. Similarly, domain scientists need to acquire the necessary computer science terminology in order to sustain the project after the collaboration is completed. 

\paragraph{Challenge(s) \& Limitation(s)}
Time pressure and unawareness of the pitfalls of terminology differences may prevent this practice from being given adequate attention. Project managers are advised to alert team members of its importance and set aside dedicated time at project begin.

\paragraph{Supportive Tools and Resources} 
While there are no specific tools that help identify and overcome terminology gaps, general collaboration tools and platforms, such as Google Docs, Wikis, GitHub pages~\cite{GitHubPages}, and Stack Overflow~\cite{stackoverflow}, have been mentioned as useful. They help develop and share glossaries, leading to improved understanding of the joint projects.

\subsection{BP6 - Understanding the Domain Problem and Developing a Project Plan}
\label{bp:domainproblem}
\begin{center}
  \fbox{\begin{minipage}{0.98\textwidth}
     Invest time in understanding the domain problem, turning possibly vague ideas into a feasible solution approach, and creating a comprehensive project plan to effectively address well-defined requirements.
  \end{minipage}}
\end{center}
\paragraph{The Practice}
Many Xpert Network participants highlighted the importance of first contact with the supported domain scientists and the approach taken to understand problems and develop solutions. In addition to awareness of the terminology gap, the relevance of investing time in understanding the goals and the functional as well as non-functional requirements was emphasized. Showing patience in the process is critical. The domain researcher needs to be helped in transforming an idea that may be initially vague into a concrete plan. Developing specific requirements for the computational application and the underlying system is essential. These requirements need to be reviewed periodically, as CDI research often evolves during the course of the project.

\paragraph{Impact}
This practice provides clarity on project objectives, challenges, and scope, establishing clear directions and focusing on critical aspects of the problem at hand. It also helps identify necessary compute and storage resources.

\paragraph{Recommended Integration Methods}
\begin{itemize}
     \item Devote significant time to understanding the domain problem and discussing/developing solutions with domain scientists. Do so at project begin as well as when research directions change. 
    \item Be sure to define and/or review goals, scope, deliverables, and milestones. Identify and challenge assumptions. Explore alternatives. Plan ahead for unexpected outcomes and contingencies.
    \item When planning, begin with high-level ideas, gradually narrowing the focus and refining details. (This point was emphasized, given different backgrounds of Xperts versus domain researchers and the terminology issue of Section~\ref{bp:terminologygap}.)
\end{itemize}

\paragraph{Difference from General Software Engineering Practices}
The collaborative relationship, with domain scientists being closely involved in the project and eventually continuing without Xpert support, as described in Section~\ref{bp:assistance}, makes this practice fundamentally different from general software engineering, where a problem is solved {\em for} the client. Furthermore, the dynamic nature of many research projects increases the importance of planning not only at project begin, but revisions when emerging research results prompt changes in project directions.

\paragraph{Challenge(s) \& Limitation(s)}
Deadline pressure can be a major impediment to this practice, as significant time is needed for the teams to get familiar with each other and with project plans. Additional time investment is needed when project directions change. Recognizing the strength that the diversity of backgrounds as well as a thorough review of goals, approaches, and assumptions can bring to the project will help overcome this challenge. This practice and its challenges go hand-in-hand with ``Overcoming the Terminology Gap,'' described in Section~\ref{bp:terminologygap}.

\paragraph{Supportive Tools and Resources} 
There are numerous project management tools available that can assist in creating and managing project plans. Some popular ones include Microsoft Project~\cite{microsoftproject}, Asana~\cite{asana}, Trello~\cite{trello}, and Jira~\cite{jira}. These tools allow for defining project scope, creating milestones, setting deadlines, and allocating human resources effectively.

\subsection{BP7 - Prioritize Functional Requirements}
\label{bp:requirements}
\begin{center}
  \fbox{\begin{minipage}{0.98\textwidth}
    Prioritize requirements based on project goals and stakeholder needs. Engage with stakeholders throughout the course of the project to gather and prioritize essential features as project needs evolve.
  \end{minipage}}
\end{center}

\paragraph{The Practice}
The dilemma of a large number of desirable features but only a short project duration can be substantial in the development of scientific software. It is important to focus on those features that are essential for the core objectives of the application and the requirements of the stakeholders. Reprioritization may be needed when research outcomes lead to new insights or stakeholder requirements change.

\paragraph{Impact}
This practice increases the efficiency of the development process and the adaptability to evolving research needs. It leads to the timely delivery of critical functionality to users, enabling early feedback for iterative improvements.

\paragraph{Recommended Integration Methods}

\begin{itemize}
    \item Gather and document all features of the to-be-developed software. Categorize the features by the source (project objectives, external stakeholders) and priority (essential, desirable, optional).
    \item Review the progressing project periodically for agreement with the set priorities. 
    \item Review, and if needed revise, priorities as research outcomes evolve. 
    \item If external stakeholders are involved, provide frequent feedback on progress and monitor requirements. Revise priorities when requirements change.
\end{itemize}
\paragraph{Difference from General Software Engineering Practices}
Functional prioritization in a research context is especially important, as there can be a large number of "nice-to-have" features of CDI applications, which however are not essential for realizing and demonstrating the core science contribution. At the same time, re-prioritization can be critical, as the research contributions may change with new results and insights.

\paragraph{Challenge(s) \& Limitation(s)}
The tendency to implement "bells and whistles" or features that are easy to realize often challenges most research software projects. Strict setting of and adherence to priorities is advised. Categorizing features from essential to optional is not always straightforward, however. For example, a seemingly optional feature of a user interface may well be the one convincing a reviewer that the developed system meets his/her bar for user-friendliness. Frequent stakeholder engagement can help address this challenge.

\paragraph{Supportive Tools and Resources} 
Tools such as Jira~\cite{jira}, and Trello~\cite{trello} can assist in capturing, organizing, prioritizing, and tracking functional requirements.

\subsection{BP8 - Issue Tracking}
\label{bp:issuetracking}
\begin{center}
  \fbox{\begin{minipage}{0.98\textwidth}
    Choose a suitable tool or method to track project requirements and bug reports, from their origin to their implementation.
  \end{minipage}}
\end{center}
\paragraph{The Practice}
Issue tracking is about systematically recording and monitoring the status of all project requirements and bug reports, from their origin to the implementation. 
Issue tracking is especially important in mid- and large-size CDI development projects, which may last several years and personnel may change. It is essential to record feature requests, implementation status, and the reasons for accepting or rejecting a request, so as to maintain a clear record and facilitate re-assessment and re-prioritization of functional requirements. This is an often overlooked practice.
 
\paragraph{Impact}
A clear record of feature requests and their implementation status enables the team to prioritize, assess, and adapt functional requirements to meet evolving project needs and possibly conflicting requirements. Issue tracking tools also prevent duplication of efforts, boost overall productivity, and reduce the risk of overlooking crucial issues, ultimately contributing to successful high-quality CDI applications. Traceability enhances transparency, accountability, and quality control throughout the development process.

\paragraph{Recommended Integration Methods}
\begin{itemize}
    \item Define a repository for collecting and tracking issues. Define the visibility and access to the repository by stakeholder groups. Be sure to consider informally communicated issues as well.
    \item Define workflows and statuses for resolving issues, documenting implementation decisions, tracking progress systematically, and reporting status updates to stakeholder groups.
    \item Periodically review unresolved issues, implementation plans, and schedules/priorities.
\end{itemize}

\paragraph{Difference from General Software Engineering Practices} 
Recall that, unlike general software engineering practices, the development of CDI applications often experiences evolving research perspectives and changing requirements. 
Issue tracking in this dynamic environment goes beyond its conventional purpose and becomes a means for recording the various paths explored during the project development. This includes documenting reasons for pivoting or persisting with certain paths. Essentially, it serves as a tool to document team decisions and illuminate the project's trajectory, adapting to the unique challenges posed by CDI applications.

\paragraph{Challenge(s) \& Limitation(s)}
Scalability is a primary challenge. At the beginning of a project, it may seem easy to track issues "in the head". However, the number of issues almost always increases beyond the point where this is feasible. The team may miss unresolved issues, fail to see conflicts between requirements, or forget to inform stakeholders. Another challenge is that clients may make bug reports or feature requests in ways most convenient for them, such as sending an email to a staff member they happen to know, where they may get lost. Without a systematic issue tracking mechanism, it is easy to get caught by such traps.

\paragraph{Supportive Tools and Resources} 
There are several methods for tracking issues in CDI development projects:
\begin{itemize}
    \item \textbf{Manual Issue Tracking:} This method involves using spreadsheets or Google documents to manually track and record the status of issues. It requires updating the tracking system by hand as issues are processed.
    \item \textbf{Visual Workflow Management:} Kanban boards~\cite{kanban} provide a visual representation of issues or tasks on a board with columns representing different stages of progress (e.g., "To Do," "In Progress," "Done"). Issues are moved across the board as they progress, providing a quick overview of their status.
    \item \textbf{Automated Issue Tracking Tools:} Dedicated Issue-Tracking Tools provide a structured and automated approach to tracking issues. These tools offer features such as creating, assigning, and updating issues, attaching files, setting priorities, and generating reports. Examples include Jira~\cite{jira}, GitHub Issues~\cite{githubissues}, Bugzilla~\cite{bugzilla}, and Trello~\cite{trello}.
\end{itemize}
The choice of tracking method depends on the size, complexity, team structure, and requirements of the project. Automated issue-tracking tools are generally recommended for larger projects or teams, as they offer more comprehensive features and scalability. Smaller projects or teams may find manual tracking or simpler tools, such as Kanban boards, sufficient.

\subsection{BP9 - Source Code Management (SCM) and Version Control}
\label{bp:sourcecode}
\begin{center}
  \fbox{\begin{minipage}{0.98\textwidth}
   Make use of source code management and version control systems to track the evolution of your software throughout its development lifecycle.
  \end{minipage}}
\end{center}
\paragraph{The Practice}
Many successful science software applications had their origin in a "toy program" written by a graduate student. It got gradually expanded by several authors, caught the attention of a wide audience, and ended up becoming an important research tool. Without version control to track the evolution of code, scripts, and data in scientific software, the history of the origin, authors, relationship of features, and specific extensions may no longer be known, making it difficult to extend further and obtain the needed documentation. 
Therefore, it is crucial for Xperts and domain scientists to learn version control methods and tools to facilitate collaborative software development, enable software roll-back to a previous, well-defined state, and start developing a new branch of the software for parallel experimentation and development. 
This practice is well known in general software engineering. Xpert Network participants emphasized it because of its importance and the likelihood of being overlooked by Xperts with non-software engineering backgrounds.

\paragraph{Impact}
Version control systems provide safety nets for experimentation, increase productivity, enhance reproducibility, and traceability of CDI applications, and thus promote scientific rigor. 

\paragraph{Recommended Integration Methods}
Xperts with computer science/engineering background are well familiar with this practice and its relevance~\cite{10.5555.1593918,fowler2001agile}. For Xperts without formal software engineering training, the inclusion of source code management and version control in the onboarding process is essential.

\paragraph{Difference from General Software Engineering Practices}
While the practice itself is the same as in general software engineering, it plays a special role in research. The intrinsically exploratory nature, with at times several research branches being tried, abandoned, merged, and down-selected, fits well the ability of SCM to fork and merge software branches. In this way, SCM tools can become important instruments for tracking, managing, and documenting research directions. This is especially important in light of the often limited time for documentation, discussed in Section~\ref{bp:documentation}.

\paragraph{Challenge(s) \& Limitation(s)}
Deadline pressure is also a major challenge in this practice, especially for Xperts not familiar with SCM tools. Convincing them of the benefits during onboarding is critical. 

\paragraph{Supportive Tools and Resources}
Among the tools are "Centralized Version Control Systems", such as Apache Subversion (SVN)~\cite{svn} and Perforce~\cite{perforce} as well as "Distributed Version Control Systems", including Git~\cite{git} and Mercurial~\cite{mercurial}.

\subsection{BP10 - Code Review}
\label{bp:codereview}
\begin{center}
  \fbox{\begin{minipage}{0.98\textwidth}
    Domain scientists review the code written by the supporting Xperts, and vice versa. Doing so will detect errors, verify common project understanding, and transfer knowledge.
  \end{minipage}}
\end{center}

\paragraph{The Practice}
Code review is a software development practice in which a second person reads the newly written code to identify errors and improve the quality of the code. 
In the context of CDI research projects, where interdisciplinary collaboration is essential, code review by the partnering Xperts and domain scientists is natural and essential. Next to the goal of error detection and code improvement, an important role of code review is to operationally refine high-level requirements and develop agreements on coding standards.

\paragraph{Impact}
While the primary goal of this practice is to improve code quality and detect bugs, code review also helps ensure a common understanding of requirements. Furthermore, it facilitates communication as well as the transfer of knowledge of the implementation, which is essential for enabling domain scientists to continue the work after the collaboration completes.

\paragraph{Recommended Integration Methods}
For basic code review methods, refer to general software engineering guidelines~\cite{fowler2001agile}. Specific recommendations for CDI projects with collaborative partnerships of Xperts and domain scientists include:
\begin{itemize}
    \item At the beginning of the collaboration, establish the timing and frequency of code reviews within the software development workflow. 
    \item Incorporate the practice into the onboarding process for new team members.
\end{itemize}

\paragraph{Difference from General Software Engineering Practices}
In CDI research, code reviews are particularly valuable because of the diverse backgrounds and expertise of Xperts and domain scientists. Mutual code review by Xperts and domain scientists leverages their combined skills and prompts a thorough examination of the code from scientific as well as technical perspectives.
Furthermore, given that research software often deals with cutting-edge technologies and novel algorithms, code reviews provide an additional safety net for these innovations.
Finally, the practice helps to pass on the expertise of an Xpert to the supported domain scientist, who will often carry forward the work once the collaboration has been completed.

\paragraph{Challenge(s) \& Limitation(s)}
Developers tend to forego code review when faced with limited resources, such as small team sizes, pressing deadlines, or lack of funding. This needs to be carefully weighed against the potential of costly efforts to recover from and fix undetected errors later in the project. 

Differing views of coding styles and practices between developer and reviewer may cost time. However, the diversity of opinions can be an opportunity as well; it may prompt fruitful discussions about best practices for software design, which may lead to guidelines that benefit the overall project.

\paragraph{Supportive Tools and Resources}
This practice can be facilitated by using specialized code review tools such as GitHub Pull Requests~\cite{githubpullrequests}, and Gerrit~\cite{gerrit}, which offer features such as code diffing, inline comments, and tracking of review comments. 

\subsection{BP11 - Test-driven Development}
\label{bp:testing}
\begin{center}
  \fbox{\begin{minipage}{0.98\textwidth}
Define the test cases that the application should pass before developing the application itself.
  \end{minipage}}
\end{center} 
\paragraph{The Practice}
Many Xpert Network participants advocated test-driven development (TDD), where test cases that will need to be passed at the end are defined before coding begins. This practice helps clarify requirements and how they will need to be demonstrated or verified at the end. TDD helps developers navigate resource constraints and prioritization, as mentioned in Section~\ref{bp:requirements}. TDD can lead to software that is lean and maintainable, by omitting functionality that is not prompted by requirements. Automated testing is highly desirable, so as to ensure that tests continue to be passed after software updates (see also Continuous Integration in Section~\ref{bp:ci}.)

\paragraph{Impact}
Comprehensive testing can detect potential issues, bugs, and inconsistencies early in the development process. Testing can also serve as a form of documentation, as it defines expected behavior. 

\paragraph{Recommended Integration Methods}
TDD is well documented in the software engineering literature~\cite{TDD2022, astels2003test}. The essence is to clarify and specify the tests that requirements need to satisfy. Complex requirements need to be broken down into components that will be tested individually. This task should be done before coding starts, rather than when it nears completion. Careful review of the tests is needed when requirements or research directions change. Developing an automated regression testing process that runs after every software update is highly recommended.

\paragraph{Difference from General Software Engineering Practices}
This practice is similar to testing in a general software engineering context. However, the intrinsically dynamic nature of research and academic priorities pose additional challenges, as described next.

\paragraph{Challenge(s) \& Limitation(s)}
Testing challenges in CDI research mirror those in general software engineering contexts -- foremost the difficulty of translating requirements into adequate tests and the time needed to do so. As mentioned above, the dynamic nature and priorities (see Section~\ref{bp:academic}) of research projects add to these difficulties. 
It is natural for research directions or user requirements to change as new scientific insights emerge. Adapting the tests that need to be satisfied by the new requirements can be both difficult and time-consuming. Academic priorities, such as a focus on demonstrating principles and concepts rather than full functionality increase these challenges. Difficult tradeoffs may need to be made between demonstrating and verifying prototype behavior versus showing adequate functionality that renders research results credible.

\paragraph{Supportive Tools and Resources}
Many tools support testing, including testing frameworks (JUnit~\cite{junit}, PyTest~\cite{pytest}),
test automation tools (Selenium~\cite{selenium}, Appium~\cite{appium}, Cypress~\cite{cypress}),
mocking and stubbing frameworks (Mockito~\cite{mockito}, Sinon~\cite{sinon}, PowerMock~\cite{powermock}),
test data generation tools (Faker~\cite{faker}, DataFactory~\cite{datafactory}, SQL Data Generator~\cite{sqldatagenerator}),
and Continuous integration and testing platforms (Jenkins~\cite{jenkins}, Travis CI~\cite{travisci}, CircleCI~\cite{circleci}).

\subsection{BP12 - Documentation}
\label{bp:documentation}
\begin{center}
  \fbox{\begin{minipage}{0.98\textwidth}
     Document your project to ensure long-term success, enable reproducibility, and obtain proper credit for your work.
  \end{minipage}}
\end{center}

\paragraph{The Practice}
Proper documentation is crucial for long-term success, reproducibility, and recognition of one's work, particularly in academic software. Documentation is essential in the joint work of Xperts with domain scientists, who will carry on the project after the collaboration is complete. However, documentation is often lacking due to time constraints and the academic reward system's focus on principles and prototypes rather than production-ready software. Navigating these differing demands is key.

\paragraph{Impact}
Proper documentation enhances reproducibility, usability, maintainability, and requirement traceability. In academic settings and CDI contexts, documentation is especially important due to frequent staff (students, postdocs) turnover, project handover at the end of collaborative assistance periods, and academic credit.

\paragraph{Recommended Implementation Methods}
\begin{itemize}
    \item Define the minimal but sufficient documentation needed for the different stakeholder groups: developers (Xperts and domain scientists involved in the development), domain scientists sustaining the project long-term, and end users. Include glossaries.
   \item Focus the descriptions on the overall project structure and workflows. Make use of in-line documentation and documentation-generation tools for code details. Be sure to include references to existing publications about the research.
\end{itemize}

\paragraph{Difference from General Software Engineering Practices} 
The need for reproducible scientific results is a key demand on documentation that is not usually present in general software engineering. This need exacerbates the problem of finding good tradeoffs between time investment in documentation versus project advancement. On the other hand, available publications about the underlying science can contribute to the documentation.

\paragraph{Challenge(s) \& Limitation(s)}
The academic reward system (Section~\ref{sec:academic-env}) and time pressure are among the primary challenges for this practice as well. Unawareness of the need for targeting different audiences (Xperts, domain scientists, end users) and a tendency to focus on details rather than high-level descriptions compound the problem. To counteract these difficulties, descriptions should focus on high-level architectures and workflows; implementation details can be explained through in-line comments combined with documentation-generation tools. The latter can also help address challenges posed by changing research directions, which can render programs and their descriptions obsolete; documentation generated from properly inserted in-line comments can more easily keep up with code modifications.

\paragraph{Supportive Tools and Resources} 
Google Docs has been mentioned as a widely useful platform for collaborative editing, including the creation of documentation. Other tools, such as Doxygen~\cite{doxygen}, Javadoc~\cite{javadoc}, Sphinx~\cite{sphinx}, and MkDocs~\cite{mkdocs}, automatically generate documentation from comments within the source code, producing documentation in formats such as HTML, PDF, or Markdown. 
Furthermore, version control systems, such as Git, are able to track and version documentation files effectively, making them readily accessible alongside code changes. 

\subsection{BP13 - Continuous Integration}
\label{bp:ci}
\begin{center}
  \fbox{\begin{minipage}{0.98\textwidth}
    Regularly incorporate new software updates into the main application version to share the latest features and identify potential issues.
  \end{minipage}}
\end{center}
\paragraph{The Practice}
Continuous Integration (CI) involves regularly merging code changes into the primary software version to make the latest updates available to other developers and users. While this practice may be well understood in industrial software development teams, Xpert Network participants reported several examples in which graduate students "sat on their codes" until graduation. Sharing the program and its features early could have benefited both end users and fellow researchers. It also could have served to identify issues with usability, compatibility with other features, miscommunicated requirements, defined test cases, and integration. Identifying such issues late in the development cycles, such as after the student has graduated and left, can be very costly.
The process of CI typically involves automated build and testing procedures (see also Section~\ref{bp:testing}), using a version control system for managing the software versions (see Section~\ref{bp:sourcecode}).
CI may be extended to include continuous deployment (CD)~\cite{CICD, improveCICD}, where successful builds are automatically deployed to production or staging environments. 

\paragraph{Impact}
Continuous integration can significantly improve team productivity by making new features available to other developers. It also reduces the cost of late-discovered issues with usability, compatibility with other features, broken requirements, failed regression tests, and incompatibilities in integration. 
Furthermore, getting accustomed to this practice can create a more iterative software development mindset, frequently incorporating feedback from both fellow developers and users into the project. This agility is particularly valuable in scientific research, where requirements and understanding of the problem domain can evolve rapidly.

\paragraph{Recommended Integration Methods}
\begin{itemize}
\item Define required integration periods as well as an integration workflow. Familiarization with supporting tools and practices for tasks such as version tracking and test generations is key. Include this material in the onboarding process.
\item Automate the CI process as much as possible, including regression testing with an evolving test suite. 
\end{itemize}

\paragraph{Difference from General Software Engineering Practices}
The importance of CI is less understood in research than in general software engineering teams. However, the dynamic nature of science increases the relevance of this practice, as changing requirements make the described issues more likely and their late discovery more costly.
CI can also prompt communication among CDI application developers in useful ways, assisting the practices outlined in Sections~\ref{bp:terminologygap} and~\ref{bp:domainproblem}.

\paragraph{Challenge(s) \& Limitation(s)}
Unawareness of the importance of CI in academic settings and the resources needed to set up and maintain the necessary infrastructure are the main impediments to this practice.

\paragraph{Supportive Tools and Resources}
\textit{Jenkins}~\cite{jenkins} is an open-source automation server for the automation of building, testing, and deployment processes. \textit{Travis CI}~\cite{travisci} offers a cloud-based CI service with easy setup and integration with GitHub. 
\textit{CircleCI}~\cite{circleci} provides a user-friendly interface and supports building, testing, and deploying applications across multiple platforms.
\textit{GitLab CI/CD}~\cite{gitlabcicd} integrates seamlessly into the GitLab platform, enabling the definition of CI/CD pipelines using configuration files. 
\textit{Bamboo}~\cite{bamboo} by Atlassian offers CI/CD capabilities and integrates with other Atlassian tools, such as Jira~\cite{jira}. \textit{TeamCity}~\cite{teamcity} by JetBrains supports the building, testing, and deployment of applications and supports various programming languages and frameworks. 
\textit{GitHub Actions}~\cite{githubactions} allow for the automation of workflows directly within GitHub repositories. 
\textit{Azure DevOps}~\cite{azuredevops} by Microsoft provides a comprehensive CI/CD platform with a range of tools and services for building, testing, and deploying applications on the Azure cloud platform. 

\subsection{BP14 - Reproducibility}
\label{bp:reproducibility}
\begin{center}
  \fbox{\begin{minipage}{0.98\textwidth}
  Enable reproducibility and transparency of your research findings by capturing and sharing the scientific method, workflow, and their associated parameters.
  \end{minipage}}
\end{center}

\paragraph{The Practice}
Reproducibility of research results refers to the ability to replicate research findings using the same methods and data that were used in the original study. This is an important requirement for all scientific research, particularly in the context of CDI research. 
By capturing the experiment workflow and its parameters (including input data, software, hardware platform, exact versions and options used, intermediate data, and results) and sharing this information through available platforms,  
researchers can increase transparency and enable others to reproduce their experiments. This practice is of growing importance~\cite{wilkinson2016fair, national2019reproducibility}, and platforms for sharing reproducibility data are becoming increasingly widespread (e.g., conferences and journals that allow for or even require reproducibility artifact descriptions.)

\paragraph{Impact}
By providing access to the data, code, and methods used in a study, reproducibility allows other researchers to independently verify and validate the results. This promotes transparency, accountability, and trust in the scientific process. Reproducibility also enables the replication and extension of studies, fosters collaboration, and knowledge sharing, and supports the advancement of scientific knowledge. Additionally, the practice promotes open science principles and encourages the sharing of research artifacts.

\paragraph{Recommended Integration Method}
Methods that promote reproducible research are well documented in the literature~\cite{wilkinson2016fair, national2019reproducibility}. They were emphasized by Xpert Network participants for their growing importance. Implementation methods cover documentation, data management, code versioning, techniques for capturing comprehensive experimental contexts, open source distribution, as well as sharing and archiving practices. 

\paragraph{Difference from General Software Engineering Practices}
Reproducibility is a key demand of scientific experimentation and hence unique to software development in research. It is rarely found in general software engineering. 

\paragraph{Challenge(s) \& Limitation(s)} 
Among the impediments to this practice are: (i) Capturing the full environmental context can be very difficult. For example, there may be settings in the operating system or architecture that the researcher is not even aware of, but which may affect performance. There may also be external dependencies on tools or data that the researcher has no control over. (ii) Architectures (both hardware and software) change rapidly and may render the exact reproduction of a computational experiment infeasible after only a few years. (iii) There is intrinsic uncertainty and often variability in the scientific observation and experiment. Their quantification is a challenge in itself. (iv) Research involving sensitive data or proprietary software and hardware may explicitly limit reproducibility.

\paragraph{Supportive Tools and Resources}
Supportive tools and resources for the practice of reproducibility in scientific research include:
\begin{itemize}
    \item Version control systems discussed in Section \ref{bp:sourcecode}, which effectively manage and share code, data, and documentation. 
    \item Containerization technologies, such as Docker~\cite{docker} or Singularity~\cite{singularity}, which package software applications and their dependencies into self-contained environments to ensure that software can run consistently across different systems, enhancing reproducibility.
    \item Platforms such as the Open Science Framework (OSF)~\cite{OSF}, osf.io, provide researchers with comprehensive tools for project management, version control, and collaborative workspaces, enabling organized and transparent research workflows. OSF is designed as a collaborative project management platform. 
    \item  Zenodo~\cite{zenodo} and figshare~\cite{figshare} are repositories that facilitate the sharing and preservation of diverse research outputs, assigning Digital Object Identifiers (DOIs) to ensure proper citation and permanent accessibility.
    \item For computational reproducibility, Code Ocean~\cite{codeocean} offers a cloud-based platform that allows researchers to encapsulate code, data, and dependencies in executable containers, promoting the transparency of computational environments. Code Ocean enables researchers to create and share "capsules," which are self-contained environments containing code, data, and dependencies. Users can execute code within the Code Ocean platform, eliminating issues related to varying local environments and configurations. 
\end{itemize}

\subsection{BP15 - Parallelization}
\label{bp:parallelization}
\begin{center}
  \fbox{\begin{minipage}{0.98\textwidth}
   Create parallel code by writing fully debugged serial code first, then parallelizing it.
  \end{minipage}}
\end{center}
\paragraph{The Practice}
Parallel execution is a fundamental strategy for gaining high performance in CDI applications and optimizing resource usage. The question of whether or not to write parallel code directly or serial code first, followed by parallelization, is decades old. While advocates of the latter point out that, by writing serial code first, one may choose algorithms that limit parallelism, the Xpert Network participants have clearly expressed a preference for that practice. One argument was that the complexity of doing both steps at once -- writing correct code and parallelizing it -- is highly complex. Another reason was that there is a scarcity of software development tools -- debuggers in particular -- for direct parallel programming. Performing the same in two distinct steps also opens the door for differential debugging methods, where an experimental program, or program section, is compared to one that is known to be correct.

\paragraph{Impact}
This practice was reported to not only save software engineering time but also reduce unpredictably long debugging times. The latter may be caused by combinations of algorithmic and parallel programming errors, which can be excessively difficult to diagnose with state-of-the-art parallel programming and error detection tools.

\paragraph{Recommended Integration Method}
This paper does not discuss the first step of writing correct serial code. We refer to the extensive software engineering literature, with Software Carpentry~\cite{wilson2006software} being a good starting point. To effectively integrate parallelization into your project, consider the following approaches:

\begin{itemize}
\item Start by profiling and analyzing the program’s performance to identify critical sections that can benefit most from parallelization.
\item Choose a method for applying parallelization. (i) Automatic parallelization, often built into compilers, is typically disabled by default due to its unpredictable effects, requiring software engineers to experiment to optimize its use. (ii) Manual parallelization is more labor-intensive and includes such techniques as annotating program sections with parallel directives or completely rewriting them in parallel. The availability of a correct, serial program is important in this process, as it can be used
to verify changes incrementally. (iii) Semi-automatic parallelization combines these approaches, utilizing automatic tools for sections amenable to automation and manual optimization for others. 
\end{itemize}

\paragraph{Difference from General Software Engineering Practices}
With almost all processors today being parallel, this practice is equally relevant for general software engineering tasks where performance matters. It is central to CDI applications, however, as the processing of large data volumes and high-performance execution are often primary concerns.

\paragraph{Challenge(s) \& Limitation(s)} 
Parallelization requires comprehending such issues as thread safety, data dependencies, load balancing, and scalability. Mastering these skills can exacerbate the challenges related to complex algorithms and systems. The skills include recognizing code that cannot be parallelized and thus needs to be replaced with new, parallel algorithms. Covering these topics in the onboarding process is crucial.

\paragraph{Supportive Tools and Resources}
\begin{itemize}
     \item Parallel programming models and libraries include OpenMP~\cite{openmp} for shared-memory parallelism, MPI~\cite{mpi} for distributed-memory parallelism, and CUDA~\cite{cuda} or OpenACC~\cite{OpenACC} for GPU programming. 
     
    \item Tools for performance analysis and debugging include a range of instruments, such as profiling tools (GNU gprof~\cite{gnugprof}, Scalasca~\cite{scalasca}, HPCToolkit~\cite{hpctoolkit}), debugging tools (\cite{allineaddt}, TotalView~\cite{totalview}),
    and performance anlysis tools (VTune~\cite{intelvtune}, TAU~\cite{tau}, NVIDIA Nsight~\cite{nvidiansight}).
       
    \item Parallelization Tools include automatic parallelizers, such as ICC (Intel C Compiler)~\cite{tian2002intel}, Rose~\cite{rose}, Cetus~\cite{cetus}, Par4all~\cite{amini2012par4all}, Pluto~\cite{bondhugula2008pluto},  Polly~\cite{grosser2011polly}, ParaWise~\cite{johnson2005parawise}, ParaGraph~\cite{bluemke2010c}, SUIF (Stanford University Intermediate Format)~\cite{wilson1994suif}, and Polaris~\cite{blume1995polaris}. There are also interactive parallelizers, such as iCetus~\cite{icetus}, SUIF explorer~\cite{liao1999suif}, and ParTool~\cite{mishra2011partool}. Many of these tools are supported by research projects.
    \end{itemize}

\subsection{BP16 - Sustainability \& Maintainability}
\label{bp:maintainability}
\begin{center}
  \fbox{\begin{minipage}{0.98\textwidth}
Consider sustainability and maintainability as core software design goals to ensure software longevity.
  \end{minipage}}
\end{center}\paragraph{The Practice}
Sustainability aims to ensure the longevity of CDI applications. This goal is rarely in the mind of academic developers, such as graduate students or post-doctoral researchers, who typically see time horizons of just a few years. However, design for sustainability can substantially reduce costs and increase long-term success. There is also increasing demand from research sponsors, who typically fund novel developments through short-term grants, but expect long-term benefits from their investments. Maintainability contributes to sustainability; at a basic level, it keeps an existing application operational, but it also aims to facilitate easy extensions and make an application tolerant of future technology changes.

\paragraph{Impact}
Next to the mentioned benefits of longevity, cost savings, and success, this practice also yields generally higher-quality CDI applications.

\paragraph{Integration Strategies}
Many of the described practices increase sustainability,
including those related to code readability, documentation, version control, continuous integration, and testing. Many common software engineering practices also contribute, such as modular software design methods and containerization techniques~\cite{IBMEngineeringSustainability2023, venters2018software}.

\paragraph{Difference from General Software Engineering Practices}
While the practice itself is the same as in general software engineering contexts, the generally shorter-term outlook of academic projects makes its consideration especially important. Next to the increasing demand for sustainability plans of funding agencies, the practice's influence on reduced long-term costs and increased success further adds to its relevance in research projects.

\paragraph{Challenge(s) \& Limitation(s)} 
Navigating the tradeoff between time investments for the long-term benefits of this practice versus immediate project needs represents one difficulty. Another challenge is to anticipate changes in technology for deciding how to best "future-proof" the application.

\paragraph{Supportive Tools and Resources}
As many of the mentioned practices contribute to sustainability and maintainability, the tools and resources listed in those sections have relevance here as well. Of particular interest are container platforms and tools, such as Docker~\cite{docker} and Kubernetes~\cite{kubernetes}. They help "future-proof" applications by packaging them with other software components they depend on, thus enhancing portability across evolving platforms.

\subsection{BP17 - User Community Engagement and Exchange}
\label{bp:communityEngagement}
\begin{center}
  \fbox{\begin{minipage}{0.98\textwidth}
  Enhance usability and adoption through frequent user feedback.
  \end{minipage}}
\end{center}

\paragraph{The Practice} 
The risk of creating application software that does not align with user expectations is significant. This often occurs because users cannot clearly articulate their needs without first engaging with the application or without being aware of the possibilities that developers and emerging technologies can bring to the table. Additionally, user requirements can differ widely among clients. In the absence of close user engagement, developers tend to implement their own interpretations of what users want, or they may focus on adding ``cool features''.
Frequent user feedback is key, especially for applications that have feature-rich user interfaces and broad audiences. Establishing effective feedback collection mechanisms is therefore a crucial element of the project plan.
This practice aligns with the principles of Continuous Integration of Section~\ref{bp:ci}. By continuously integrating feedback into the development process, software is more likely to meet user needs effectively and evolve in tune with evolving demands.

\paragraph{Impact} 
The adoption of this practice helps ensure that the final application aligns closely with user expectations, as a result mitigating the potential for costly post-deployment modifications. Additionally, this approach promotes community engagement, leading to a sense of ownership and endorsement among users. The practice not only drives wider adoption but also encourages active user participation in the development lifecycle, which can yield a more refined and user-centric product.

\paragraph{Recommended Integration Methods}
\begin{itemize}
\item In the project plan, define the intended user communities and the methods for collecting their requirements and feedback. These could take the form of online community commenting tools, product demonstrations, workshops, presentations at conferences and professional meetings, surveys, and direct contacts.
\item Take advantage of the tight interaction of Xperts and domain scientists to clarify and refine requirements.
\item Define periodic feedback reviews, potentially in conjunction with Continuous Integration cycles, to analyze received community comments and incorporate them into the evolving project plan.
\end{itemize}

\paragraph{Difference from General Software Engineering Practices}
This practice is essential in industry software product developments but often overlooked in academic settings.
However, due to the dynamic nature of science, with evolving research directions and changing requirements, community engagement is vital.

\paragraph{Challenge(s) \& Limitation(s)}
The time investment needed for this practice is, again, a primary obstacle. It needs to be weighed against the possible cost of broken or misunderstood requirements. 
Another issue is that users often provide feedback through informal channels that differ from those predefined by developers, such as phone calls, casual remarks during meetings, or as side notes in emails that cover a variety of topics. 
It is important to capture these pieces of feedback and integrate them into the central feedback repository.
A further issue is the dynamic nature of science. As in several other practices, it is a significant challenge and further increases the importance of keeping tight links with the user community.

A factor that helps overcome these challenges is the
collaborative environment of Xperts and domain scientists
(as detailed in Section~\ref{bp:assistance}), which forms a natural context for collecting feedback and refining requirements. Another natural form of community engagement results from presentations at conferences, professional meetings, and peer networking, which are part of academic culture. These venues can serve as a valuable source of feedback.

\paragraph{Supportive Tools and Resources}
Feedback mechanism tools, such as UserVoice~\cite{uservoice}, SurveyMonkey~\cite{surveymonkey}, and Google Forms~\cite{googleforms}, are instrumental in collecting feedback, preferences, and opinions from users. 

\section{Evaluation}
\label{evaluation}
To assess the efficacy of best practices, we utilized three distinct evaluation methods: (i) researcher surveys, (ii) a case study of the Atom Project, and (iii) expert reviews. Each method evaluates the usability and impact of the practices from a distinct perspective -— general CDI researchers, a combination of Xperts and domain scientists involved in a specific research project, and Xperts assisting in the development of CDI applications, respectively. We also collected data on the experience levels of the survey participants.
For each evaluation, we developed survey forms tailored to the specific needs of the participants and the study. All of these forms included a brief description of the practices under review prior to soliciting feedback from the participants. The metrics evaluated and the questions asked varied across the different forms to suit each distinct methodological approach.
Sections~\ref{sec:ressurvey} through \ref{sec:xpsurvey} describe each survey, including objectives, participants, and an analysis of the results.

\subsection{Researcher Surveys} \label{sec:ressurvey}
The researcher surveys gauge the research community's familiarity with identified practices and assess their impact on CDI applications. This evaluation was conducted when the initial 15 practices were identified by the Xpert Network effort.
The survey sought feedback from CDI researchers across various disciplines utilizing HPC resources at the University of Delaware. Participants possessed domain knowledge and some experience in CDI application development.

The survey process briefly introduced researchers to the 15 identified best practices. The participants were then asked to assess the practices on two key metrics: (i) {\em Experience} evaluates the research community's experience in adopting the practices. (ii) {\em Impact} measures the effects of implementing practices, helping identify those that yield promising results and significantly benefit the field. While ``No Experience'' indicates that the practice has not been used, "Low Impact" and "High Impact" indicate the corresponding experience with the practice.

\paragraph{Survey Data and Findings}
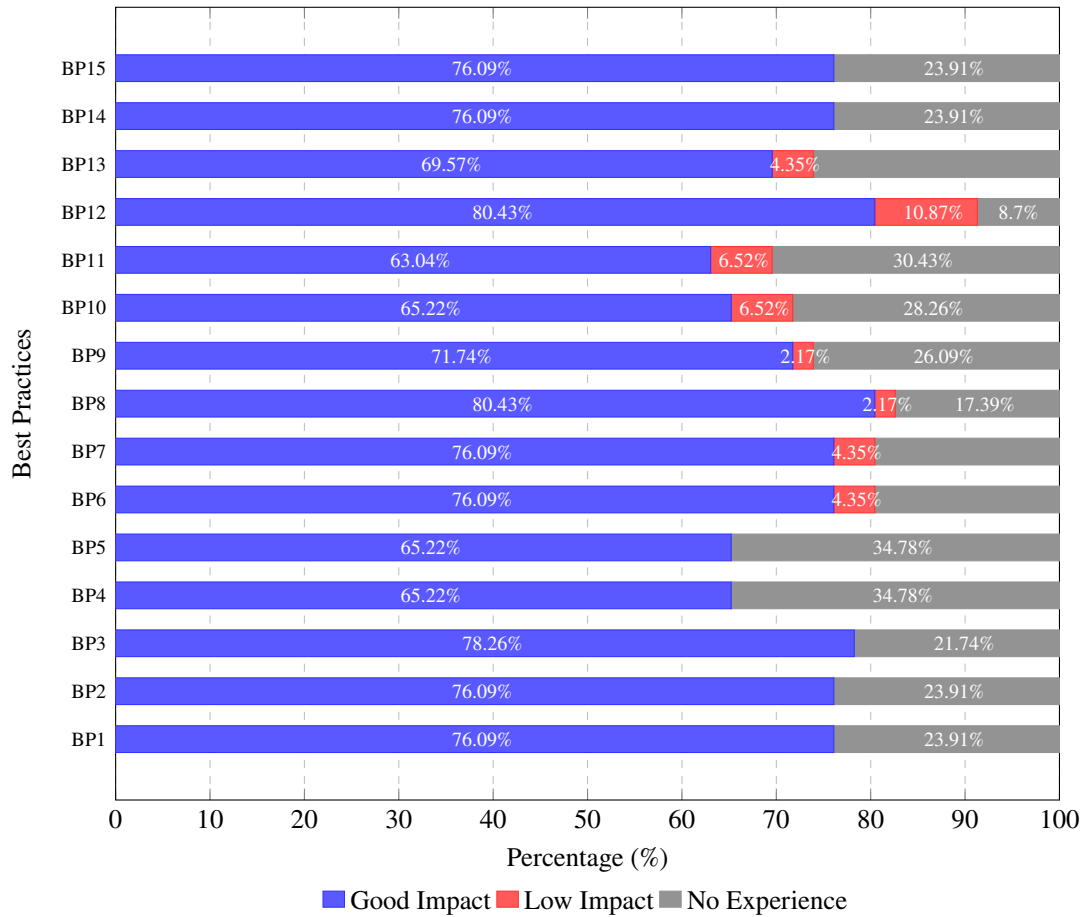
\begin{figure}
\centering

\pgfplotsset{width=8cm,compat=1.15}
\begin{tikzpicture}
\begin{axis}[
    xbar stacked,       
    width=14cm,
    height=12cm,
    enlarge y limits={abs=0.8cm},
    xmin=0,
    xmax=100,
    xlabel={Percentage (\%)},
    ylabel={Best Practices},
    ytick={1,...,15},
    yticklabels={
        BP1, BP2, BP3, BP4, BP5, BP6, BP7, BP8, BP9, BP10, BP11, BP12, BP13, BP14, BP15
    },
    legend style={
        at={(0.5,-0.1)},
        anchor=north,
        legend columns=-1,
        draw=none
    },
    yticklabel style={
        align=right,
        text width=0.75cm,
        anchor=east,
        font=\footnotesize, 
    },
    grid style={dashed, gray!50},
    xmajorgrids=true,
    nodes near coords={
        \pgfmathprintnumber{\pgfplotspointmeta}\%
    },
    every node near coord/.append style={
        font=\footnotesize,
        anchor=west,
        rotate=0,
        text width=0.6cm,
        align=center,
        color=white,
        xshift=-12pt,
    },
    enlarge x limits=0.0,
]
\addplot[blue!80,fill=blue!65, nodes near coords={ \pgfmathprintnumber{\pgfplotspointmeta}\%}] coordinates {
    (76.09, 1)
    (76.09, 2)
    (78.26, 3)
    (65.22, 4)
    (65.22, 5)
    (76.09, 6)
    (76.09, 7)
    (80.43, 8)
    (71.74, 9)
    (65.22, 10)
    (63.04, 11)
    (80.43, 12)
    (69.57, 13)
    (76.09, 14)
    (76.09, 15)
};
\addplot[red!80,fill=red!65, nodes near coords={ \pgfmathprintnumber{\pgfplotspointmeta}\%
}] coordinates {
    (0.00, 1)
    (0.00, 2)
    (0.00, 3)
    (0.00, 4)
    (0.00, 5)
    (4.35, 6)
    (4.35, 7)
    (2.17, 8)
    (2.17, 9)
    (6.52, 10)
    (6.52, 11)
    (10.87, 12)
    (4.35, 13)
    (0.00, 14)
    (0.00, 15)
};
\addplot[gray!85,fill=gray!85, nodes near coords={ \pgfmathprintnumber{\pgfplotspointmeta}\%}] coordinates {
    (23.91, 1)
    (23.91, 2)
    (21.74, 3)
    (34.78, 4)
    (34.78, 5)
    (19.57, 6)
    (19.57, 7)
    (17.39, 8)
    (26.09, 9)
    (28.26, 10)
    (30.43, 11)
    (8.70, 12)
    (26.09, 13)
    (23.91, 14)
    (23.91, 15)
};
\legend{Good Impact, Low Impact, No Experience}
\end{axis}
\end{tikzpicture}
\caption{ - Researcher Survey Results (46 participants)}
\label{ResearcherSurvey}
\justify\footnotesize
BP1 - Onboarding Xperts from Diverse Backgrounds; 
BP2 - Understanding the Academic Environment; 
BP3 - Developing a Breadth of Skills for Effective Handling of Projects; 
BP4 - Collaborative Assistance Between Xperts and Domain scientists; 
BP5 - Overcoming the Terminology Gap; 
BP6 - Understanding the Domain Problem; 
BP7 - Prioritize Functional Requirements; 
BP8 - Issue Tracking; 
BP9 - Source Code Management; 
BP10 - Code Review; 
BP11 - Software Testing; 
BP12 - Documentation; 
BP13 - Continuous Integration; 
BP14 - Reproducibility; 
BP15 - Parallelization; 
\end{figure}
\normalsize

Figure \ref{ResearcherSurvey} on page \pageref{ResearcherSurvey} presents data from the researcher surveys. We identify the following common trends:

\begin{itemize}
    \item All practices show "Good Impact" (63.04\% to 80.43\%) on CDI software development. 
    \item Correspondingly there is a low occurrence of "Low Impact" across all practices (0\% to 10.87\%), indicating the positive effectiveness of the practices. 
    \item There is substantial variation in familiarity with the practices, with 8.7\% to 34.78\% reporting no prior experience with the practices. This result suggests that additional educational outreach and training may increase adoption. 
\end{itemize}

The survey data also point to potential areas for improvement:
\begin{itemize}
    \item Low Impacts: Despite the overall positive feedback, a small proportion of ``Low Impact'' responses, particularly in Code Review (6.52\%), Software Testing (6.52\%), and Documentation (10.87\%), points to challenges that may be mitigated through targeted training and better resource allocation for the application of these practices.     
    \item Limited Adoption: The limited application of certain practices, notably ``BP4 - Collaborative Assistance Between Xperts and Domain Scientists'' and ``BP5 - Overcoming the Terminology Gap Between Computer and Domain Sciences'' (both at 34.78\%), as well as ``BP11 - Software Testing'' (30.43\%), indicate a lack of exposure to interdisciplinary collaboration or early-stage research engagement. These findings suggest the importance of fostering interdisciplinary projects and integrating these practices more effectively across research phases.         
\end{itemize}

\subsection{Case Study Survey of the Atom Project}
\label{sec:CaseStudySurvey}
To further evaluate the best practices, we applied them in a case study referred to as the Atom Project~\cite{Atom, AtomGateways21, AtomGateways22}, collecting feedback from the involved participants. Recall from Section~\ref{sec:atom} that this study aimed to gauge the effectiveness and completeness of the practices in a science project in which the authors were directly involved; the study identified two additional, relevant practices (BP16 - Sustainability \& Maintainability and BP17 - User Community Engagement and Exchange).

\paragraph{Survey Explanation}
The case study survey evaluated all 17 practices identified by the Xpert Network and the Atom project, targeting computational experts and developers involved in the Atom project. We collected three metrics: (i) experience with each practice, (ii) the practice's impact, and (iii) its usability. Additionally, we asked the participants about methods and tools they found useful for employing the practices as well as about experiences with challenges, outcomes, benefits, and limitations of the practices.

\paragraph{Survey Data and Findings}

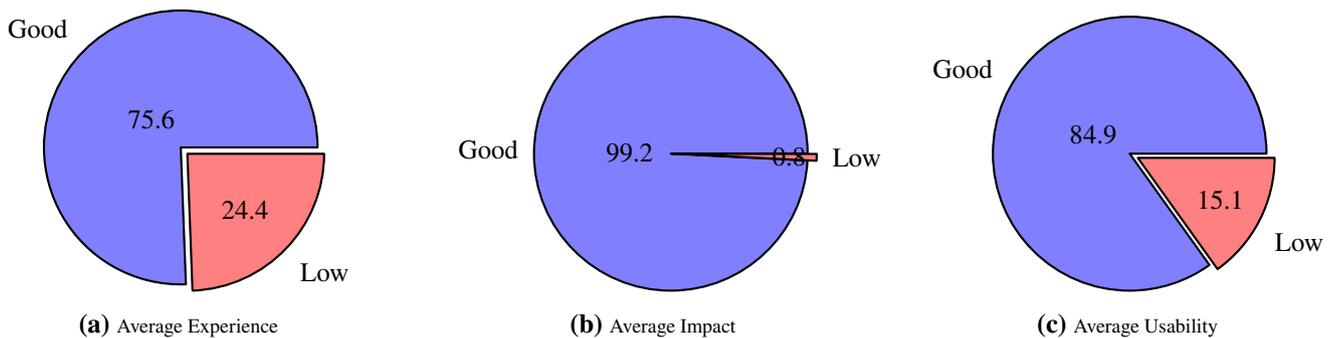
\begin{figure}[ht]
  \centering
  \begin{subfigure}[b]{0.3\textwidth}
    \centering
    \begin{tikzpicture}[scale=0.6]
      \pie[sum=auto, explode=0.1, color={blue!50, red!50}]
      {75.6/Good, 24.4/Low}
    \end{tikzpicture}
     \caption[{\footnotesize Average Experience}]{\footnotesize Average Experience}
  \end{subfigure}\hfill%
  \begin{subfigure}[b]{0.3\textwidth}
    \centering
    \begin{tikzpicture}[scale=0.6]
       \pie[sum=auto, explode=0.1, color={blue!50, red!50}]
      {99.2/Good, 0.8/Low}
    \end{tikzpicture}
     \caption[{\footnotesize Average Impact}]{\footnotesize Average Impact}
  \end{subfigure}\hfill%
  \begin{subfigure}[b]{0.3\textwidth}
    \centering
    \begin{tikzpicture}[scale=0.6]
      \pie[sum=auto, explode=0.1, color={blue!50, red!50}]
      {84.9/Good, 15.1/Low}
    \end{tikzpicture}
    \caption[{\footnotesize Average Usability}]{\footnotesize Average Usability}
  \end{subfigure}
    \caption{- Average Percentage Ratings for All Practices in the Case Study Survey by Metric Overview}
    \label{fig:AtomSurveyAveragePie}

\end{figure}

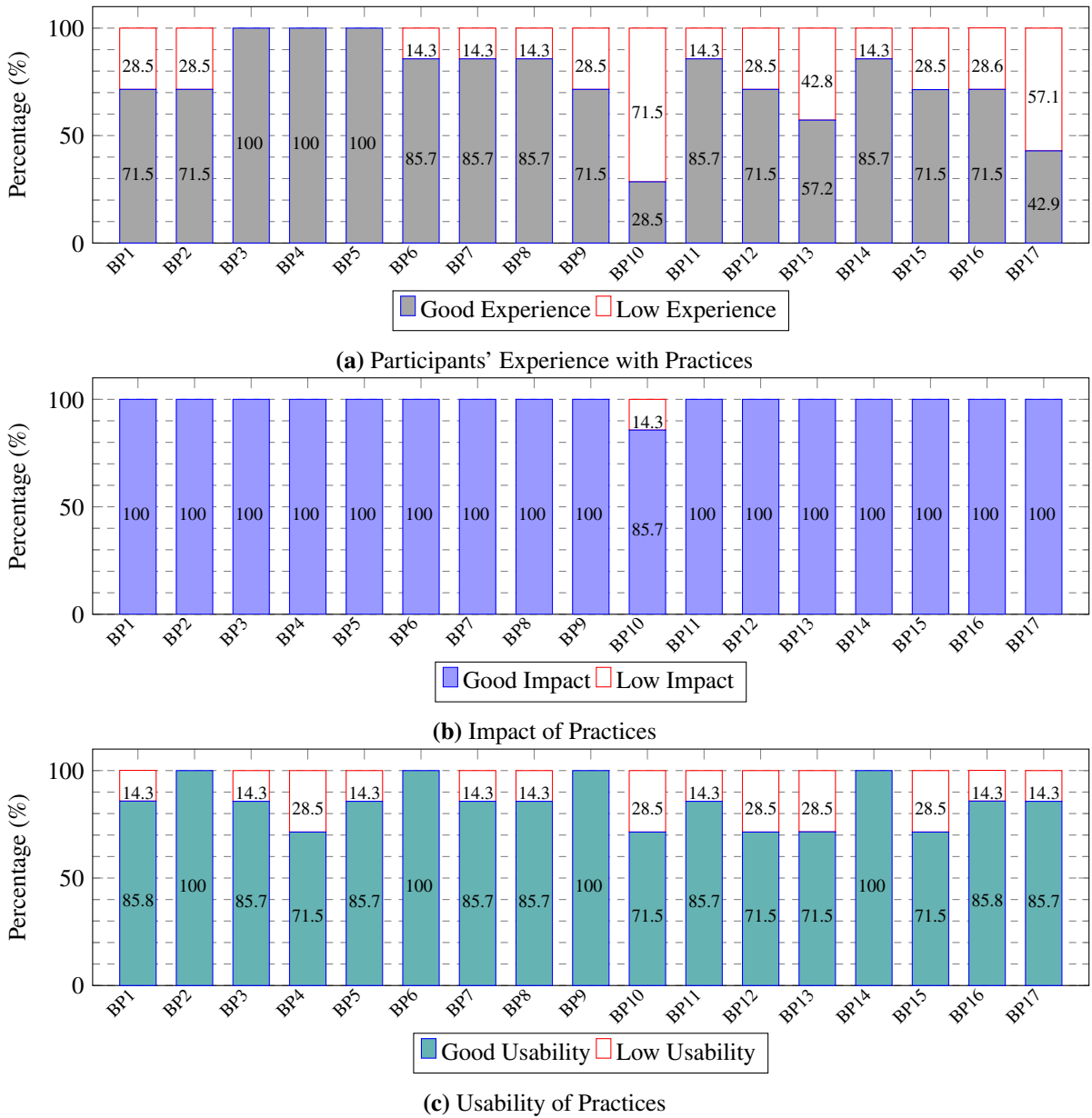
\begin{figure}[htp]
    \centering
    \begin{subfigure}[b]{1\textwidth}
\centering
\begin{tikzpicture}
\begin{axis}[
    ybar stacked,   
    bar width=15pt,
    width=16cm,
    height=5cm,
    ymin=0, ymax=110,
    enlarge x limits=0.05,
    legend style={at={(0.5,-0.20)}, anchor=north,legend columns=-1},
    ylabel={Percentage (\%)},
    ymajorgrids=true,
    yminorgrids=true,
    minor y tick num=4,
    grid style={dashed, gray},
    symbolic x coords={
        BP1, Gap1, 
        BP2, Gap1,
        BP3, Gap1,
        BP4, Gap1,
        BP5, Gap1,
        BP6, Gap1,
        BP7, Gap1,
        BP8, Gap1,
        BP9, Gap1,
        BP10, Gap1,
        BP11, Gap1,
        BP12, Gap1,
        BP13, Gap1,
        BP14, Gap1,
        BP15, Gap1,
        BP16, Gap1,
        BP17
    },
    xtick=data,
    x tick label style={font=\footnotesize, rotate=45, anchor=east},
    nodes near coords align={center},
    every node near coord/.append style={
        font=\footnotesize,
        anchor=west,
        rotate=0,
        text width=0.6cm,
        xshift=-10pt,
        yshift=-3pt,
        color=black
    },
    point meta=explicit symbolic,
    xticklabel style={
        check for zero/.code={
            \pgfmathfloatifflags{\tick}{0}{\pgfkeys{/pgf/number format/.cd,fixed,precision=0}}{},
            \ifdim\tick pt=0pt
                \pgfmathparse{int(mod(\tick,3))}
                \ifnum\pgfmathresult=0
                    \pgfplotsset{xticklabel=\empty}
                \fi
            \fi
        },
    }
]

\addplot+[fill=gray!70, nodes near coords] coordinates {
    (BP1,71.5) [71.5]
    (BP2,71.5) [71.5]
    (BP3,100.0) [100]
    (BP4,100.0) [100]
    (BP5,100.0) [100]
    (BP6,85.7) [85.7]
    (BP7,85.7) [85.7]
    (BP8,85.7) [85.7]
    (BP9,71.5) [71.5]
    (BP10,28.5) [28.5]
    (BP11,85.7) [85.7]
    (BP12,71.5) [71.5]
    (BP13,57.2) [57.2]
    (BP14,85.7) [85.7]
    (BP15,71.4) [71.5]
    (BP16,71.5) [71.5]
    (BP17,42.9) [42.9]
};
\addlegendentry{Good Experience}

\addplot+[fill=gray!0, nodes near coords] coordinates {
    (BP1,28.5) [28.5]
    (BP2,28.5) [28.5]
    (BP3,0) 
    (BP4,0) 
    (BP5,0) 
    (BP6,14.3) [14.3]
    (BP7,14.3) [14.3]
    (BP8,14.3) [14.3]
    (BP9,28.5) [28.5]
    (BP10,71.5) [71.5]
    (BP11,14.3) [14.3]
    (BP12,28.5) [28.5]
    (BP13,42.8) [42.8]
    (BP14,14.3) [14.3]
    (BP15,28.6) [28.5]
    (BP16,28.6) [28.6]
    (BP17,57.1) [57.1]
};
\addlegendentry{Low Experience}

\end{axis}
\end{tikzpicture}
\caption{Participants' Experience with Practices }
\label{fig:AtomExperience}
\end{subfigure}
\begin{subfigure}[b]{1\textwidth}
\centering
\begin{tikzpicture}
\begin{axis}[
    ybar stacked,   
    bar width=15pt,
    width=16cm,
    height=5cm,
    ymin=0, ymax=110, 
    enlarge x limits=0.05, 
    legend style={at={(0.5,-0.20)}, anchor=north,legend columns=-1},
    ylabel={Percentage (\%)},
    ymajorgrids=true,
    yminorgrids=true,
    minor y tick num=4,
    grid style={dashed, gray},
    symbolic x coords={
        BP1,  Gap1, 
        BP2,  Gap1,
        BP3, Gap1,
        BP4,  Gap1,
        BP5,  Gap1,
        BP6, Gap1,
        BP7,  Gap1,
        BP8,  Gap1,
        BP9,  Gap1,
        BP10,  Gap1,
        BP11,  Gap1,
        BP12,  Gap1,
        BP13,  Gap1,
        BP14,  Gap1,
        BP15,  Gap1,
        BP16,  Gap1,
        BP17
    },
    xtick=data,
    x tick label style={font=\footnotesize, rotate=45, anchor=east},
    nodes near coords align={center},
    every node near coord/.append style={
        font=\footnotesize,
        anchor=west,
        rotate=0,
        text width=0.6cm,
        xshift=-10pt,
        yshift=-3pt,
        color=black
    },
    point meta=explicit symbolic,
    xticklabel style={
        check for zero/.code={
            \pgfmathfloatifflags{\tick}{0}{\pgfkeys{/pgf/number format/.cd,fixed,precision=0}}{},
            \ifdim\tick pt=0pt
                \pgfmathparse{int(mod(\tick,3))}
                \ifnum\pgfmathresult=0
                    \pgfplotsset{xticklabel=\empty}
                \fi
            \fi
        },
    }
]

\addplot+[fill=blue!40, nodes near coords] coordinates {
    (BP1,100.0) [100]
    (BP2,100.0) [100]
    (BP3,100.0) [100]
    (BP4,100.0) [100]
    (BP5,100.0) [100]
    (BP6,100.0) [100]
    (BP7,100.0) [100]
    (BP8,100.0) [100]
    (BP9,100.0) [100]
    (BP10,85.7) [85.7]
    (BP11,100.0) [100]
    (BP12,100.0) [100]
    (BP13,100.0) [100]
    (BP14,100.0) [100]
    (BP15,100.0) [100]
    (BP16,100.0) [100]
    (BP17,100.0) [100]
};
\addlegendentry{Good Impact}

\addplot+[fill=blue!0, nodes near coords] coordinates {
    (BP1,0.0) 
    (BP2,0.0) 
    (BP3,0.0) 
    (BP4,0.0) 
    (BP5,0.0) 
    (BP6,0.0) 
    (BP7,0.0) 
    (BP8,0.0) 
    (BP9,0.0) 
    (BP10,14.3) [14.3]
    (BP11,0.0) 
    (BP12,0.0) 
    (BP13,0.0) 
    (BP14,0.0) 
    (BP15,0.0) 
    (BP16,0.0) 
    (BP17,0.0) 
};
\addlegendentry{Low Impact}

\end{axis}
\end{tikzpicture}
\caption{Impact of Practices}
\label{fig:AtomImpact}
\end{subfigure}
\begin{subfigure}[b]{1\textwidth}
\centering
\begin{tikzpicture}
\begin{axis}[
    ybar stacked,   
    bar width=15pt,
    width=16cm,
    height=5cm,
    ymin=0, ymax=110, 
    enlarge x limits=0.05, 
    legend style={at={(0.5,-0.20)}, anchor=north,legend columns=-1},
    ylabel={Percentage (\%)},
    ymajorgrids=true,
    yminorgrids=true,
    minor y tick num=4,
    grid style={dashed, gray},
    symbolic x coords={
        BP1, Gap1, 
        BP2, Gap1,
        BP3, Gap1,
        BP4, Gap1,
        BP5, Gap1,
        BP6, Gap1,
        BP7, Gap1,
        BP8, Gap1,
        BP9, Gap1,
        BP10, Gap1,
        BP11, Gap1,
        BP12, Gap1,
        BP13, Gap1,
        BP14, Gap1,
        BP15, Gap1,
        BP16, Gap1,
        BP17
    },
    xtick=data,
    x tick label style={font=\footnotesize, rotate=45, anchor=east},
    nodes near coords align={center},
    every node near coord/.append style={
        font=\footnotesize,
        anchor=west,
        rotate=0,
        text width=0.6cm,
        xshift=-10pt,
        yshift=-3pt,
        color=black
    },
    point meta=explicit symbolic,
    xticklabel style={
        check for zero/.code={
            \pgfmathfloatifflags{\tick}{0}{\pgfkeys{/pgf/number format/.cd,fixed,precision=0}}{},
            \ifdim\tick pt=0pt
                \pgfmathparse{int(mod(\tick,3))}
                \ifnum\pgfmathresult=0
                    \pgfplotsset{xticklabel=\empty}
                \fi
            \fi
        },
    }
]

\addplot+[fill=teal!60, nodes near coords] coordinates {
    (BP1,85.8) [85.8]
    (BP2,100.0) [100]
    (BP3,85.7) [85.7]
    (BP4,71.4) [71.5]
    (BP5,85.7) [85.7]
    (BP6,100.0) [100]
    (BP7,85.7) [85.7]
    (BP8,85.7) [85.7]
    (BP9,100.0) [100]
    (BP10,71.4) [71.5]
    (BP11,85.7) [85.7]
    (BP12,71.4) [71.5]
    (BP13,71.5) [71.5]
    (BP14,100.0) [100]
    (BP15,71.4) [71.5]
    (BP16,85.8) [85.8]
    (BP17,85.7) [85.7]
};
\addlegendentry{Good Usability}

\addplot+[fill=teal!0, nodes near coords] coordinates {
    (BP1,14.3) [14.3]
    (BP2,0.0) 
    (BP3,14.3) [14.3]
    (BP4,28.6) [28.5]
    (BP5,14.3) [14.3]
    (BP6,0.0) 
    (BP7,14.3) [14.3]
    (BP8,14.3) [14.3]
    (BP9,0.0) 
    (BP10,28.6) [28.5]
    (BP11,14.3) [14.3]
    (BP12,28.6) [28.5]
    (BP13,28.5) [28.5]
    (BP14,0.0) 
    (BP15,28.6) [28.5]
    (BP16,14.3) [14.3]
    (BP17,14.3) [14.3]
};
\addlegendentry{Low Usability}

\end{axis}
\end{tikzpicture}
\caption{Usability of Practices}
\label{fig:AtomUsability}
\end{subfigure}

\caption{ - Collected Data from the Case Study Surveys}
\label{fig:vertical_caseStudy}
\justify\footnotesize
BP1 - Onboarding Xperts from Diverse Backgrounds; 
BP2 - Understanding the Academic Environment; 
BP3 - Developing a Breadth of Skills for Effective Handling of Projects; 
BP4 - Collaborative Assistance Between Xperts and Domain scientists; 
BP5 - Overcoming the Terminology Gap; 
BP6 - Understanding the Domain Problem; 
BP7 - Prioritize Functional Requirements; 
BP8 - Issue Tracking; 
BP9 - Source Code Management; 
BP10 - Code Review; 
BP11 - Software Testing; 
BP12 - Documentation; 
BP13 - Continuous Integration; 
BP14 - Reproducibility; 
BP15 - Parallelization; 
BP16 - Maintainability \& Sustainability; 
BP17 - User Community Engagement and Exchange; 
\end{figure}

Figure~\ref{fig:AtomSurveyAveragePie} on page~\pageref{fig:AtomSurveyAveragePie} shows the average percentage ratings for all practices across each survey metric. 
The overall average ``Good'' rating for each metric significantly exceeds the ``Low'' rating portion. This response indicates the substantial expertise of the participants and their recognition of the practices' benefits and ease of integration into the development workflow.
Figure~\ref{fig:vertical_caseStudy} on page~\pageref{fig:vertical_caseStudy} allows 
a more in-depth analysis of the data.

\begin{itemize}
    \item Figure~\ref{fig:AtomExperience} on page~\pageref{fig:AtomExperience} shows that most practices have experience ratings exceeding 50\%, suggesting that survey participants exhibit a ``Good'' level of familiarity and competency with the practices.
    However, certain practices, such as BP10 and BP17, show an experience level below 50\%. This disparity may be attributed to the diverse backgrounds and skill sets of the domain scientists and Xperts participating in the survey.
    \item Figure~\ref{fig:AtomImpact} on page~\pageref{fig:AtomImpact} illustrates a predominantly positive impact, with "Good impact" ratings reaching 100\% for nearly all practices. Practice 10 -- Code Review -- is an exception, with a "Good impact" of 85.7\%. We attribute this lower rating to lesser awareness of the benefits of code review. 
    \item Figure~\ref{fig:AtomUsability} on page~\pageref{fig:AtomUsability} indicates a high level of "Good usability" for practices (above 71.5\%), suggesting that the practices have been relatively easy to integrate into the development process and have been well-utilized by the participants.
\end{itemize}

The following practices have low usability above the 15\% average; each of these practices reported low usability at 28.5\%:

\begin{itemize}
    \item \textbf{Practice 4 - Collaborative Assistance Between Xperts and Domain Scientists:} While short-term collaborations are adequate for many projects, Atom Project participants have highlighted the limitations of this approach for the project. They preferred long-term partnerships, diverging from the advice provided by Xperts who reported effective short-term collaborations. 
    \item \textbf{Practice 10 - Code Review:} Participants expressed difficulties in integrating code review into the development process, especially when it falls outside their domain expertise. Understanding each other's code can be challenging, hindering the smooth integration of code review. 
    However, gaining a good understanding of each other's code is crucial, especially when the project will eventually be continued by the domain scientists once the collaboration concludes. 
    \item \textbf{Practice 12 - Documentation:} The lower usability of this practice was linked to the fact that, although it was relatively straightforward to keep developer documentation up to date by adding comments within the code, maintaining separate documentation tailored specifically for domain scientists, in this case physicists, required additional effort. Ensuring that the documentation met the needs and understanding of domain scientists demanded more resources and attention. 
    \item \textbf{Practice 13 - Continuous Integration:} The low usability rate of this practice within the project can be attributed to multiple factors, including the challenges of developing comprehensive test cases, automating the testing process, and setting up the necessary infrastructure, tools, and automated workflows.
    \item \textbf{Practice 15 - Parallelization:} The reported low usability, as noted by some participants, is attributed to the requirement for specialized skills beyond the expertise of domain scientists in effectively implementing parallelization techniques in CDI projects. 
\end{itemize}

\subsection{Expert Reviews}
\label{sec:xpsurvey}
This survey targeted CDI Xperts, possessing advanced knowledge, experience, and practical insights in the field of CDI research. We selected these experts from among the most active participants in the Xpert Network. 
The survey included all 17 practices of Section~\ref{BestPractices}.
The survey began with a brief introduction of each practice, followed by evaluations from the experts across three metrics: (i) their experience level with the practice, (ii) the practice's realized impact, and (iii) usability, specifically focusing on ease and efficiency of implementation in real-life scenarios.
Respondents were asked to rate each metric as either ``Good'' or ``Low''. Additionally, they were encouraged to share the methods and tools they used in applying the practices. This information was presented in the tools paragraphs of Section~\ref{BestPractices}.

\begin{figure}[ht]
  \centering

  \begin{subfigure}[b]{0.3\textwidth}
    \centering
    \begin{tikzpicture}[scale=0.6]
      \pie[sum=auto, explode=0.1, color={blue!50, red!50}]
      {85.8/Good, 14.2/Low}
    \end{tikzpicture}
     \caption[{\footnotesize Average Experience}]{\footnotesize Average Experience}
  \end{subfigure}\hfill%
  \begin{subfigure}[b]{0.3\textwidth}
    \centering
    \begin{tikzpicture}[scale=0.6]
      \pie[sum=auto, explode=0.1, color={blue!50, red!50}]
      {96.7/Good, 3.4/Low}
    \end{tikzpicture}
     \caption[{\footnotesize Average Impact}]{\footnotesize Average Impact}
  \end{subfigure}\hfill%
  \begin{subfigure}[b]{0.3\textwidth}
    \centering
    \begin{tikzpicture}[scale=0.6]
      \pie[sum=auto, explode=0.1, color={blue!50, red!50}]
      {89.1/Good, 10.9/Low}
    \end{tikzpicture}
     \caption[{\footnotesize Average Usability}]{\footnotesize Average Usability}
  \end{subfigure}
    \caption{- Average Percentage Ratings for All Practices in the Expert Review Survey by Metric Overview}
    \label{fig:ExpertReviewAveragePie}
\end{figure}
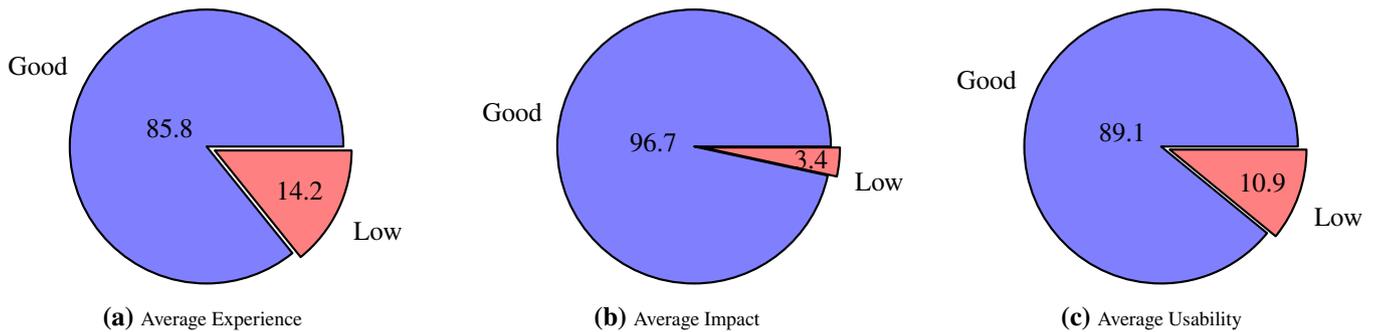

\paragraph{Survey Data and Findings}
Figure~\ref{fig:ExpertReviewAveragePie} on page~\pageref{fig:ExpertReviewAveragePie} displays an average percentage rating of all practices for each metric, depicted through pie charts. The sum of the ``Good'' rating in each metric significantly outweighs the ``Low'' rating. This shows that the participants possess substantial expertise in the subject matter, assess the practices as having positive impacts, and find them usable and well-integrated with the development process. A more detailed analysis of each metric follows.

\paragraph{Evaluation of Experience}

\begin{figure}[htp]
\centering
\begin{subfigure}[b]{1\textwidth}
\centering
\begin{tikzpicture}
\begin{axis}[
    ybar stacked,   
    bar width=15pt,
    width=16cm,
    height=5cm,
    ymin=0, ymax=110, 
    enlarge x limits=0.05, 
    legend style={at={(0.5,-0.20)}, anchor=north,legend columns=-1},
    ylabel={Percentage (\%)},
    ymajorgrids=true, 
    yminorgrids=true, 
    minor y tick num=4, 
    grid style={dashed, gray}, 
    symbolic x coords={
    BP1, Gap1, 
    BP2, Gap1,
    BP3, Gap1,
    BP4, Gap1,
    BP5, Gap1,
    BP6, Gap1,
    BP7, Gap1,
    BP8, Gap1,
    BP9, Gap1,
    BP10, Gap1,
    BP11, Gap1,
    BP12, Gap1,
    BP13, Gap1,
    BP14, Gap1,
    BP15, Gap1,
    BP16, Gap1,
    BP17
    }, 
    xtick=data,
    x tick label style={font=\footnotesize}, 
    x tick label style={rotate=45,anchor=east},
    nodes near coords align={center},
    every node near coord/.append style={
        font=\footnotesize,
        anchor=west,
        rotate=0,
        text width=0.6cm,
        xshift=-10pt,
        yshift=-3pt,
        color=black
    },
    point meta=explicit symbolic, 
    xticklabel style={ 
        check for zero/.code={
            \pgfmathfloatifflags{\tick}{0}{\pgfkeys{/pgf/number format/.cd,fixed,precision=0}}{},
            \ifdim\tick pt=0pt
                \pgfmathparse{int(mod(\tick,3))}
                \ifnum\pgfmathresult=0
                    \pgfplotsset{xticklabel=\empty}
                \fi
            \fi
        },
    }
]
\addplot+[fill=gray!70, nodes near coords ] coordinates {
(BP1,100.0)  [100]
(BP2,85.8)  [85.8]
(BP3,78.6)  [78.6]
(BP4,92.9)  [92.9]
(BP5,92.9)  [92.9]
(BP6,92.9)  [92.9]
(BP7,92.9)  [92.9]
(BP8,85.8)  [85.8]
(BP9,78.6)  [78.6]
(BP10,85.8)  [85.8]
(BP11,78.6)  [78.6]
(BP12,85.8) [85.8]
(BP13,64.3)  [64.3]
(BP14,92.9)  [92.9]
(BP15,85.8) [85.8]
(BP16,78.6)  [78.6]
(BP17,85.8)  [85.8]
};
\addlegendentry{ Good Exp}
\addplot+[fill=gray!0, nodes near coords] coordinates {
(BP1,0) 
(BP2,14.2) [14.2]
(BP3,21.4) [21.4]
(BP4,7.1) [7.1]
(BP5,7.1) [7.1]
(BP6,7.1) [7.1]
(BP7,7.1) [7.1]
(BP8,14.2) [14.2]
(BP9,21.4) [21.4]
(BP10,14.2) [14.2]
(BP11,21.4) [21.4]
(BP12,14.2) [14.2]
(BP13,35.7) [35.7]
(BP14,7.1) [7.1]
(BP15,14.2) [14.2]
(BP16,21.4) [21.4]
(BP17,14.2) [14.2]
};
\addlegendentry{ Low Exp}

\end{axis}
\end{tikzpicture}
\caption{Experience of Xperts with Practices}
\label{fig:ExpertReviewExperience}
\end{subfigure}
\begin{subfigure}[b]{1\textwidth}
\centering
\begin{tikzpicture}
\begin{axis}[
    ybar stacked,   
    bar width=15pt,
    width=16cm,
    height=5cm,
    ymin=0, ymax=110, 
    enlarge x limits=0.05, 
    legend style={at={(0.5,-0.20)}, anchor=north,legend columns=-1},
    ylabel={Percentage (\%)},
    ymajorgrids=true, 
    yminorgrids=true, 
    minor y tick num=4, 
    grid style={dashed, gray}, 
    symbolic x coords={
    BP1,  Gap1, 
    BP2,  Gap1,
    BP3, Gap1,
    BP4,  Gap1,
    BP5,  Gap1,
    BP6, Gap1,
    BP7,  Gap1,
    BP8,  Gap1,
    BP9,  Gap1,
    BP10,  Gap1,
    BP11,  Gap1,
    BP12,  Gap1,
    BP13,  Gap1,
    BP14,  Gap1,
    BP15,  Gap1,
    BP16,  Gap1,
    BP17
    }, 
    xtick=data,
    x tick label style={font=\footnotesize}, 
    x tick label style={rotate=45,anchor=east},
    nodes near coords align={center},
    every node near coord/.append style={
        font=\footnotesize,
        anchor=west,
        rotate=0,
        text width=0.6cm,
        xshift=-10pt,
        yshift=-3pt,
        color=black
    },
    point meta=explicit symbolic, 
    xticklabel style={ 
        check for zero/.code={
            \pgfmathfloatifflags{\tick}{0}{\pgfkeys{/pgf/number format/.cd,fixed,precision=0}}{},
            \ifdim\tick pt=0pt
                \pgfmathparse{int(mod(\tick,3))}
                \ifnum\pgfmathresult=0
                    \pgfplotsset{xticklabel=\empty}
                \fi
            \fi
        },
    }
]
\addplot+[fill=blue!40, nodes near coords] coordinates {
(BP1,100.0) [100]
(BP2,100.0) [100]
(BP3,92.9) [92.9]
(BP4,100.0) [100]
(BP5,92.9) [92.9]
(BP6,92.9) [92.9]
(BP7,92.9) [92.9]
(BP8,100.0) [100]
(BP9,92.9) [92.9]
(BP10,92.9) [92.9]
(BP11,92.9) [92.9]
(BP12,100.0) [100]
(BP13,100.0) [100]
(BP14,92.9) [92.9]
(BP15,100.0) [100]
(BP16,100.0) [100]
(BP17,100.0) [100]
};
\addlegendentry{Good Impact}
\addplot+[fill=blue!0, nodes near coords] coordinates {
(BP1,0.0) 
(BP2,0.0) 
(BP3,7.1) [7.1]
(BP4,0.0) 
(BP5,7.1) [7.1]
(BP6,7.1) [7.1]
(BP7,7.1) [7.1]
(BP8,0.0) 
(BP9,7.1) [7.1]
(BP10,7.1) [7.1]
(BP11,7.1) [7.1]
(BP12,0.0) 
(BP13,0.0) 
(BP14,7.1) [7.1]
(BP15,0.0) 
(BP16,0.0) 
(BP17,0.0) 
};
\addlegendentry{Low Impact}

\end{axis}
\end{tikzpicture}
\caption{Impact of Best Practices}
\label{fig:ExpertReviewImpact}
\end{subfigure}
\begin{subfigure}[b]{1\textwidth}
\centering
\begin{tikzpicture}
\begin{axis}[
    ybar stacked,   
    bar width=15pt,
    width=16cm,
    height=5cm,
    ymin=0, ymax=110, 
    enlarge x limits=0.05, 
    legend style={at={(0.5,-0.20)}, anchor=north,legend columns=-1},
    ylabel={Percentage (\%)},
    ymajorgrids=true, 
    yminorgrids=true, 
    minor y tick num=4, 
    grid style={dashed, gray}, 
    symbolic x coords={
    BP1, Gap1, 
    BP2, Gap1,
    BP3, Gap1,
    BP4, Gap1,
    BP5, Gap1,
    BP6, Gap1,
    BP7, Gap1,
    BP8, Gap1,
    BP9, Gap1,
    BP10, Gap1,
    BP11, Gap1,
    BP12, Gap1,
    BP13, Gap1,
    BP14, Gap1,
    BP15, Gap1,
    BP16, Gap1,
    BP17
    }, 
    xtick=data,
    x tick label style={font=\footnotesize}, 
    x tick label style={rotate=45,anchor=east},
    nodes near coords align={center},
    every node near coord/.append style={
        font=\footnotesize,
        anchor=west,
        rotate=0,
        text width=0.6cm,
        xshift=-10pt,
        yshift=-3pt,
        color=black
    },
    point meta=explicit symbolic, 
    xticklabel style={ 
        check for zero/.code={
            \pgfmathfloatifflags{\tick}{0}{\pgfkeys{/pgf/number format/.cd,fixed,precision=0}}{},
            \ifdim\tick pt=0pt
                \pgfmathparse{int(mod(\tick,3))}
                \ifnum\pgfmathresult=0
                    \pgfplotsset{xticklabel=\empty}
                \fi
            \fi
        },
    }
]

\addplot+[fill=teal!60, nodes near coords] coordinates {
(BP1,100.0) [100]
(BP2,85.8) [85.8]
(BP3,92.9) [92.9]
(BP4,92.9) [92.9]
(BP5,100.0) [100]
(BP6,92.9) [92.9]
(BP7,92.9) [92.9]
(BP8,92.9) [92.9]
(BP9,100.0) [100]
(BP10,71.4) [71.4]
(BP11,85.8) [85.8]
(BP12,92.9) [92.9]
(BP13,78.6) [78.6]
(BP14,85.8) [85.8]
(BP15,71.4) [71.4]
(BP16,92.9) [92.9]
(BP17,85.8) [85.8]
};
\addlegendentry{Good Usability}
\addplot+[fill=teal!0, nodes near coords] coordinates {
(BP1,0.0)
(BP2,14.2) [14.2]
(BP3,7.1) [7.1]
(BP4,7.1) [7.1]
(BP5,0.0) 
(BP6,7.1) [7.1]
(BP7,7.1) [7.1 ]
(BP8,7.1) [7.1 ]
(BP9,0.0)
(BP10,28.6) [28.6]
(BP11,14.2) [14.2]
(BP12,7.1) [7.1]
(BP13,21.4) [21.4]
(BP14,14.2) [14.2]
(BP15,28.6) [28.6]
(BP16,7.1) [7.1]
(BP17,14.2) [14.2]
};
\addlegendentry{Low Usability}
\end{axis}
\end{tikzpicture}
\caption{Usability of Best Practices}
\label{fig:ExpertReviewUsability}

\end{subfigure}

\caption{ - Collected Data from Expert Reviews Surveys}
\label{fig:vertical_expertReviews}
\justify\footnotesize
BP1 - Onboarding Xperts from Diverse Backgrounds; 
BP2 - Understanding the Academic Environment; 
BP3 - Developing a Breadth of Skills for Effective Handling of Projects; 
BP4 - Collaborative Assistance Between Xperts and Domain scientists; 
BP5 - Overcoming the Terminology Gap; 
BP6 - Understanding the Domain Problem; 
BP7 - Prioritize Functional Requirements; 
BP8 - Issue Tracking; 
BP9 - Source Code Management; 
BP10 - Code Review; 
BP11 - Software Testing; 
BP12 - Documentation; 
BP13 - Continuous Integration; 
BP14 - Reproducibility; 
BP15 - Parallelization; 
BP16 - Maintainability \& Sustainability; 
BP17 - User Community Engagement and Exchange; 
\end{figure}
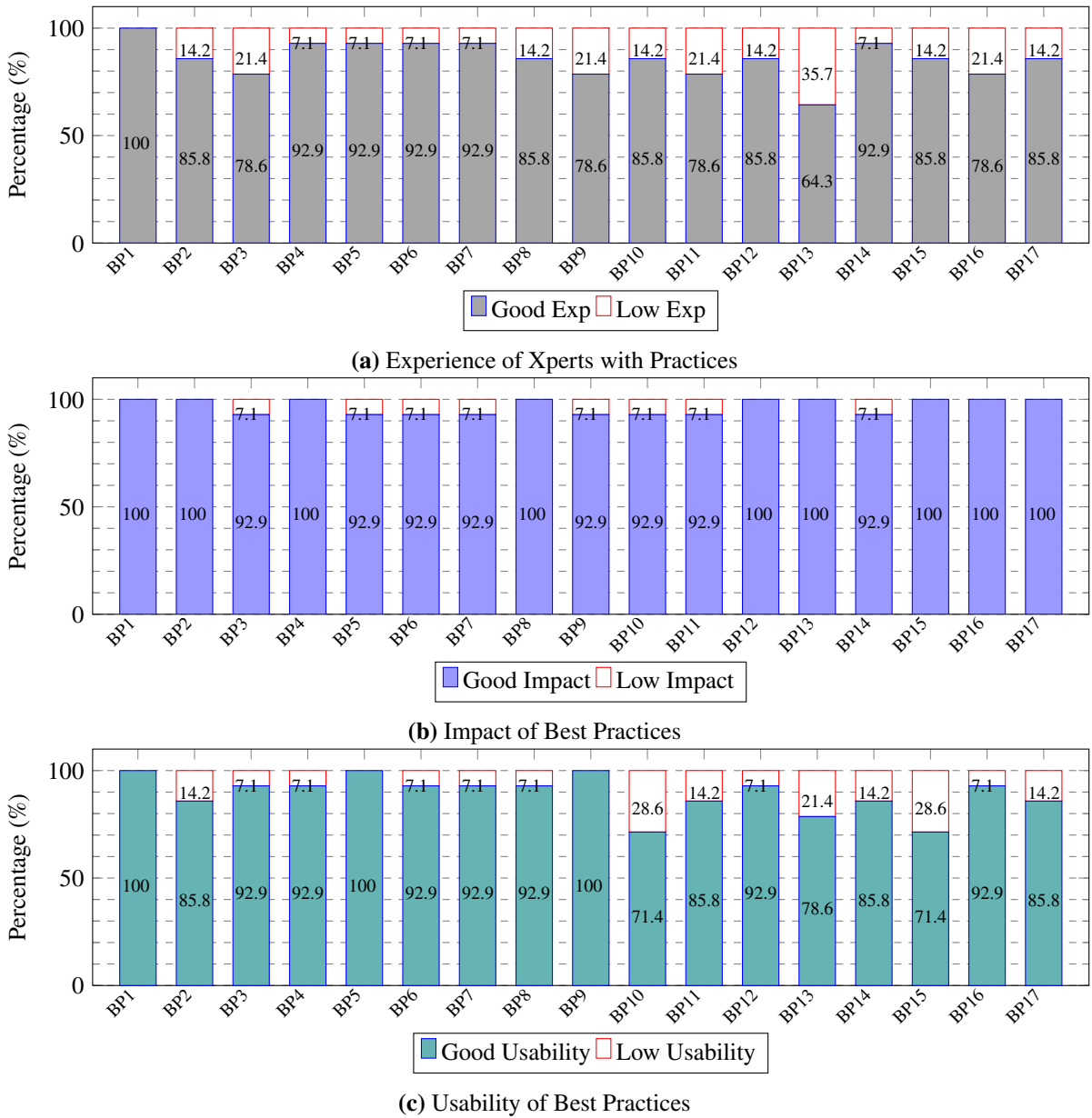

\begin{itemize}
    \item Figure~\ref{fig:ExpertReviewExperience} on page~\pageref{fig:ExpertReviewExperience} shows that, overall, the respondents have a moderate to high level of ``Good'' experience in applying the practices. Variations in experience levels may stem from the experts' diverse backgrounds, the range of project goals and sizes, and the different stages at which these experts were involved in their respective projects.
    \item Figure~\ref{fig:ExpertReviewImpact} on page~\pageref{fig:ExpertReviewImpact} shows a high positive impact, with ``Good'' exceeding 92\% for each practice. This implies that, with effective implementation, these practices hold the promise of delivering substantial positive impacts and results.
    \item Figure~\ref{fig:ExpertReviewUsability} on page~\pageref{fig:ExpertReviewUsability} shows a positive usability evaluation, with ``Good'' usability ratings consistently above 71\%. Certain practices exhibit lower usability due to factors such as resource limitations, and specific application challenges. 
\end{itemize}

The following practices had lower usability ratings. Contributory factors to these ratings were derived from supplementary comments provided by the participants:
\begin{itemize}
    \item \textbf{Practice 15- Parallelization \& Practice 10- Code Review:} show considerable low usability ratings (28.6\%), due to the complexity and expertise required. 
   
    \item \textbf{Practice 13- Continuous Integration:} faces usability challenges (21.4\% low usability), possibly due to resource and training gaps.
    
    \item \textbf{Practice 11- Software Testing,  Practice 14- Reproducibility \& Practice 17- User Community Engagement and Exchange:} encounter usability obstacles (14.2\% low usability), linked to insufficient resources or time constraints, insufficient guidance, and engagement difficulties.
    
    \item \textbf{Practice 2- Understanding the Academic Environment:} also displays usability concerns (14.2\% low usability), reflecting the complexity of navigating diverse academic settings.    
\end{itemize}

These insights highlight areas for focused improvement to enhance the adoption and effectiveness of the practices. Addressing the identified challenges, through enhanced training, better resource allocation, and clearer guidance, can significantly improve usability and overall impact. Further strategies to address these challenges are discussed in Section~\ref{BestPractices}.

\subsection{Comparing the Results of the Three Surveys}

\begin{figure}[h]
\centering
\pgfplotsset{width=8cm,compat=1.15}
\begin{subfigure}[h]{\textwidth}
\centering
\begin{tikzpicture}
\begin{axis}[
    xbar stacked,       
    width=12cm,
    height=5cm,
    enlarge y limits={abs=0.8cm},
    xmin=0,
    xmax=100,
    xlabel={Percentage},
    ylabel={Surveys on Impact},
    ytick={1,...,15},
    yticklabels={
        {Researcher Survey},
        {Expert Review Survey},
        {Case Study Survey}
    },
    legend style={
        at={(0.5,-0.30)},
        anchor=north,
        legend columns=-1,
    },
    x tick label style={font=\footnotesize},
    yticklabel style={
        align=right,
        text width=3.4cm,
        anchor=east,
        font=\footnotesize, 
    },
    grid style={dashed, gray!50},
    xmajorgrids=true,
    nodes near coords={
        \pgfmathprintnumber{\pgfplotspointmeta}\%
    },
    every node near coord/.append style={
        font=\footnotesize,
        anchor=west,
        rotate=0,
        text width=0.6cm,
        align=center,
        color=white,
        xshift=-12pt,
    },
    enlarge x limits=0.0,
]

\addplot[blue!80,fill=blue!80, nodes near coords={ \pgfmathprintnumber{\pgfplotspointmeta}\%}] coordinates {
    (73.0, 1)
    (96.6, 2)
    (99.2, 3)     
};

\addplot[red!80,fill=red!80, nodes near coords={ \pgfmathprintnumber{\pgfplotspointmeta}\%}] coordinates {
    (2.7, 1)
    (3.4, 2)
    (0.8, 3)
};
\addplot[gray!90,fill=gray!90, nodes near coords={ 
\pgfmathprintnumber{\pgfplotspointmeta}\%}] coordinates {
    (24.2, 1)
    (0, 2)
    (0, 3)  
};
\legend{Good Impact, Low Impact, No Experience}
\end{axis}
\end{tikzpicture}
\caption{ Comparing the Averages of Good and Low Impact of Practices: Expert Review, Case Study, and Researcher Survey}
\label{comparingImpact}
\end{subfigure}

\begin{subfigure}[h]{\textwidth}
\centering
\begin{tikzpicture}
\begin{axis}[
    xbar stacked,       
    width=12cm,
    height=4cm,
    enlarge y limits={abs=0.8cm},
    xmin=0,
    xmax=100,
    xlabel={Percentage},
    ylabel={Surveys on Usability},
    ytick={1,...,15},
    yticklabels={
        {Expert Review Survey},
        {Case Study Survey}
    },
    legend style={
        at={(0.5,-0.40)},
        anchor=north,
        legend columns=-1,
    },
    x tick label style={font=\footnotesize},
    yticklabel style={
        align=right,
        text width=3.4cm,
        anchor=east,
        font=\footnotesize, 
    },
    grid style={dashed, gray!50},
    xmajorgrids=true,
    nodes near coords={
        \pgfmathprintnumber{\pgfplotspointmeta}\%
    },
    every node near coord/.append style={
        font=\footnotesize,
        anchor=west,
        rotate=0,
        text width=0.6cm,
        align=center,
        color=white,
        xshift=-12pt,
    },
    enlarge x limits=0.0,
]
\addplot[blue!80,fill=blue!80, nodes near coords={ \pgfmathprintnumber{\pgfplotspointmeta}\%}] coordinates {
    (89.1, 1)
    (84.9, 2)  
};
\addplot[red!80,fill=red!80, nodes near coords={ \pgfmathprintnumber{\pgfplotspointmeta}\%}] coordinates {
        (10.9, 1)
        (15.1, 2)
};

\legend{Good Usability, Low Usability}
\end{axis}
\end{tikzpicture}
\caption{Comparing the Averages of Good and Low Usability of Practices: Expert Review Survey, and Case Study Survey}
\label{comparingUsability}
\end{subfigure}
\caption{ - Comparing the Averages of Impact and Usability of Practices}
\label{compareAllSurveys}
\end{figure}
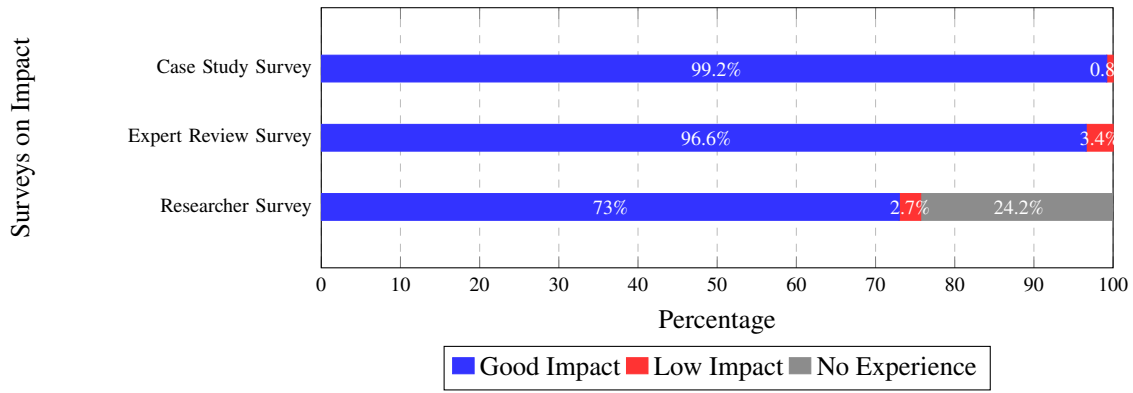
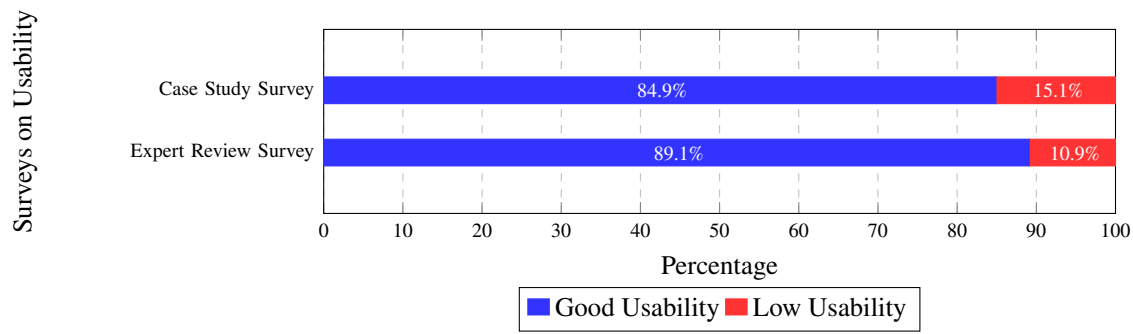

\paragraph{Impact Assessment} 
Figure~\ref{comparingImpact} on page \pageref{comparingImpact} compares the three surveys, showing the average of ``Good Impact'', ``Low Impact'', and ``No Experience'' for all practices.
``No Experience'' is reported for practices that researchers have not applied in their projects. While participants in the case study had experience implementing the practices, and participants in the expert review, who are computational experts, apply the practices daily, none reported having no experience with the practices. By contrast, the researcher survey covers a broader community, including respondents with limited experience in these practices.

All surveys consistently show high percentages of average ``Good Impact'' for the practices. only one practice -- BP10: Code Review --  shows lower impact across all surveys, as shown in Figure~\ref{ResearcherSurvey} on page ~\pageref{ResearcherSurvey}, Figure ~\ref{fig:AtomImpact} on page~\pageref{fig:AtomImpact}, and Figure ~\ref{fig:ExpertReviewImpact} on page ~\pageref{fig:ExpertReviewImpact}. One reason for this result is a lack of clear instructions on what constitutes good code and how code reviews should be conducted.

\paragraph{Usability Evaluation}
Usability assessments were conducted via the expert review and case study surveys (the researcher survey omitted this question for brevity, aiming to increase the response rates.) 
Figure~\ref{comparingUsability} shows the average reported ``Good Usability'' and ``Low Usability''. There is high overall ``Good Usability'' across practices, with a slightly higher value in the expert review survey. This result is likely due to the participants' expertise, which facilitates the smoother integration of practices into their development workflows.

In Figure~\ref{fig:AtomUsability} on page \pageref{fig:AtomUsability} and Figure~\ref{fig:ExpertReviewUsability} on page \pageref{fig:ExpertReviewUsability}, three practices show lower usability in both surveys, despite their significant impact: BP10 - Code Review, BP13 - Continuous Integration, and BP15 - Parallelization. We attribute these results to the following reasons:
\begin{itemize}
    \item Code Review Training: Users and limited training on effective code review strategies, including what constitutes high-quality code and how to identify areas for improvement.
    \item Continuous Integration Tools: While there are many tools supporting the practice, there is a lack of overall environment, including for test case generation. 
    \item Parallelization Knowledge: Parallelization skills are complex and often require extensive training.
\end{itemize}

Despite the participants' diverse experiences and perspectives, Figure~\ref{compareAllSurveys} shows consistent positive impact and usability of the practices.

\section{Related work}
\label{relatedWork}

{\bf General research software engineering efforts:} A multitude of initiatives and organizations have contributed to advancing software engineering in general, which include the development of best practices:\\
{\em Software Carpentry}~\cite{wilson2006software} is a widely recognized non-profit organization that offers workshops and online resources to teach researchers foundational software development skills. Their focus is on providing practical training on tools, programming languages, and software development practices relevant to scientific application development. Software Carpentry emphasizes reproducibility, version control, data management, and automated testing as key aspects of scientific software development.\\
{\em The Turing Way}~\cite{arnold2019turing} is an open-source community-driven project that provides a handbook on best practices in reproducible data science. It offers practical guidance and resources for researchers and software developers covering reproducible research. The Turing Way places a strong emphasis on open science, inclusivity, and fostering a collaborative research culture.\\
{\em Better Scientific Software (BSSw)}~\cite{bssw} is an initiative aimed at improving the quality and sustainability of scientific software. BSSw focuses on fostering community-driven discussions, sharing expertise, and advocating for better software practices. Their community-driven articles encompass various subjects, such as software engineering techniques, performance optimization, software licensing, and software citation. 
\\
The {\em Research Software Engineering (RSE)} community~\cite{cohen2020four} brings together professionals who are specialized in developing software for research purposes. This community focuses on advocating for the recognition and career development of research software engineers, sharing expertise, and promoting best practices in research software development. RSE publications often discuss topics such as software sustainability, collaborative development, and the role of research software engineers in interdisciplinary research. 
\\
{\em ACCESS -- Advanced Cyberinfrastructure Coordination Ecosystem: Services \& Support}~\cite{ACCESS} Builds upon the foundation laid by XSEDE~\cite{towns2014xsede}. It focuses on providing researchers with advanced cyberinfrastructure resources, services, and support to enhance the capabilities of scientific communities. Through ACCESS, researchers gain access to high-performance computing, data management, visualization tools, and networking infrastructure. ACCESS actively collaborates with researchers across diverse domains to understand their computational needs, enabling the development and deployment of customized cyberinfrastructure solutions. By bridging the gap between researchers and advanced computing resources, ACCESS empowers scientific communities to engage in large-scale simulations, data analysis, and modeling, leading to accelerated scientific discovery and fostering innovation. 
\\
The {\em Molecular Sciences Software Institute (MolSSI)}~\cite{MOLSII} tackles software challenges in molecular sciences, offering resources and training to improve scientific workflows' productivity and reproducibility. It focuses on advancing the development, maintenance, and sustainability of tools in molecular simulations and quantum chemistry. By offering resources, training, and community support, MolSII fosters best practices in software engineering, data management, and software citation, thus improving the robustness and impact of molecular sciences research. 
\\[2mm]\indent While these efforts share the common goal of improving scientific software development practices, each has its unique focus and approach. Software Carpentry and The Turing Way concentrate on providing practical training and resources for researchers to acquire essential software development skills and adopt best practices. Better Scientific Software focuses on community engagement, sharing expertise, and advocating for better software practices in the scientific research community.
The RSE community focuses on professionalizing the role of research software engineers and promoting collaboration between disciplines. ACCESS provides cyberinfrastructure resources, services, and support for researchers conducting CDI research. MolSSI addresses software challenges in molecular sciences, offering resources and training to enhance productivity and reproducibility. By contrast, this paper, while collaborating with these initiatives, aims to develop a comprehensive set of guidelines tailored for CDI application development. The guidelines include strategies for implementation, supportive tools, and insights into overcoming potential challenges, evaluated by a diverse pool of participants. 
\\[3mm]
{\bf Specific contributions towards best practices:} Wilson et al.~\cite{wilson2014best} address common challenges in scientific computing and propose eight best practices to improve the reliability of scientific software. They emphasize version control, testing, documentation, pair programming, and code review. 
The same authors introduce simple and practical steps that can improve the efficiency and reliability of scientific software~\cite{wilson2017good}. The steps represent a minimum set of tools and techniques that researchers should consider adopting. 
Both contributions are based on the collective experience of researchers involved in Software Carpentry and Data Carpentry. There has been no evaluation of the recommended practices.

Heroux et al. propose ten practices to improve computational science and engineering (CSE) software~\cite{5069157}. They address the challenges and constraints faced by CSE software developers and aim to enhance the software quality, reliability, and maintainability. The practices cover requirements gathering, version control, testing, documentation, and collaborative development. They were derived from the Trilinos project~\cite{trilinos}. There is also no evaluation of the practices provided. 

Our contribution stands out from other initiatives in that it delivers guidelines specifically crafted for Xperts who assist domain scientists in their Computational and Data Intensive (CDI) research projects. In contrast to the above-mentioned efforts, which concentrate on providing guidelines for CDI application developers, our approach highlights the importance of forming collaboration between domain scientists and Xperts, which is similar to the approach taken by the US-RSE~\cite{USRSE}
organization. Our study includes an evaluation of the practices, which is absent in all prior contributions.

Our work also discusses challenges, limitations, and implementation methods. Furthermore, we mention tools and resources that support practices. By offering a resource that addresses both technical and practical aspects, our aim is to empower Xperts to effectively navigate the complexities of CDI applications.

\section{Conclusions and Future Directions}
\label{conclusion}

Through the collective efforts of the Xpert Network and the Atom project, we have identified and evaluated a set of best practices for developing computational and data-intensive (CDI) applications. These guidelines aim to assist CDI support professionals, referred to as Xperts or Research Software Engineers, but also serve computational scientists in general. We have examined the impacts, limitations, and challenges associated with these practices, as well as recommended tools that support their effective application.

We have evaluated the practices through three distinct methods tailored to participants with varying levels of experience and perspectives. The results confirm both high impact and usability in CDI application development projects. A few of the practices are less familiar to general users; we have discussed methods to increase their adoption.

We consider this paper a live document that needs to evolve together with new technology and CDI applications. By continuing to engage a large user community to refine and extend the practices, we aim to maintain a useful guide that can be put in the hands of Xperts assisting CDI domain researchers and, in this way, help push the frontiers of computational and data science.

\section*{Acknowledgments}
This work was supported in part by the National Science Foundation under Awards No. OAC-1931339 and OAC-2209639.

\printbibliography[notkeyword={tool}, title={References}]
\printbibliography[keyword={tool}, title={References for Tools}]

\clearpage

\section*{Author Biography}
\begin{biography}

{\includegraphics[width=56pt,height=56pt]{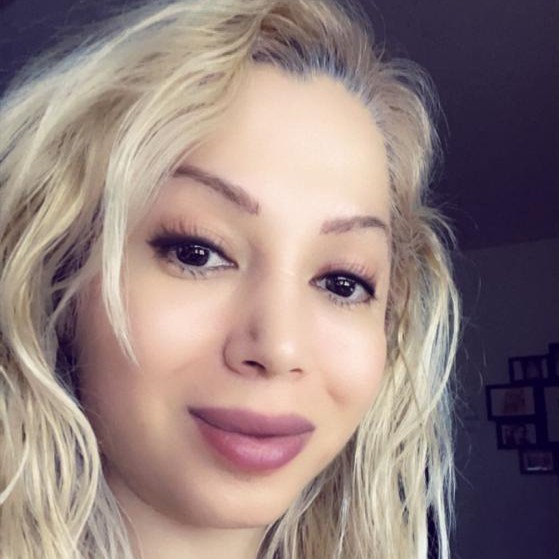}}
\noindent{\textbf{Parinaz Barakhshan.} is a Ph.D. candidate and Research Assistant at the University of Delaware, Department of Electrical and Computer Engineering. Her research interests include optimizing compilers and performance evaluation for high-performance computing systems. Contact her at parinazb@udel.edu.}

{\includegraphics[width=56pt,height=56pt]{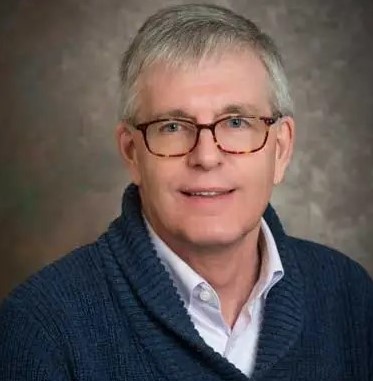}}\\{\textbf{Rudolf Eigenmann.} is a Professor at the Department of Electrical and Computer Engineering, University of Delaware, USA. His research interests include optimizing compilers, programming methodologies, tools, and performance evaluation for high-performance computing, as well as the design of cyberinfrastructure. Contact him at eigenman@udel.edu.}

\end{biography}
\end{document}